\documentclass[%
preprint,
 amsmath,amssymb,
]{revtex4-1}

\usepackage{bm}
\usepackage{color}
\usepackage{graphicx,xspace,epsfig,float,multirow,subfigure,tabularx}

\begin{document}

\title{A quantitative description of Nernst effect in high-temperature superconductors}

\author{Rong Li}
\author{Zhen-Su She}%
 \email{she@pku.edu.cn}
\affiliation{%
State Key Laboratory for Turbulence and Complex Systems,  College of Engineering, Peking University, Beijing 100871, China}%

\date{\today}

\begin{abstract}
A quantitative  vortex-fluid model for flux-flow resistivity $\rho$ and Nernst signal $e_N$ in
high-temperature superconductors (HTSC) is proposed. Two kinds of vortices, magnetic and thermal, are
considered, and the damping viscosity $\eta$ is modeled by extending the Bardeen-Stephen model to include
the contributions of flux pinning at low temperature and in weak magnetic fields, and vortex-vortex
collisions in strong magnetic fields. Remarkably accurate descriptions for both Nernst signal of six
samples and flux flow resistivity are achieved over a wide range of temperature $T$ and magnetic field $B$. A discrepancy of three orders of magnitude between data and Anderson's model of Nernst signal is pointed out and revised using experimental values of $\eta$ from magnetoresistance. Furthermore, a two-step procedure is developed to reliably extract, from the Nernst signal, a set of physical parameters characterizing the vortex dynamics, which yields predictions of local superfluid density $n_s$, the Kosterlitz coefficient $b$ of thermal vortices, and upper critical field and temperature. Application of the model and systematic measurement of relevant physical quantities from Nernst signal in other HTSC samples are discussed.
\end{abstract}

\pacs{Valid PACS appear here}
\maketitle


\section{Introduction}\label{intro}

It is well known that transport properties are powerful tools to address the pseudogap problem in HTSC \cite{Timusk1999,Tallon2001}. For example, the out-of-plane magnetoresistance $\rho_c$ has been used to
determine the onset temperature of pseudogap $T^*$ \cite{Usui2014}. A more remarkable property is the
so-called Nernst effect, in which a conductor exhibits a transverse detectable voltage when magnetic field is applied along \textbf{\^{z}} direction perpendicular to the direction of temperature gradient. Nernst signal
is usually small in simple metal. But for a superconductor composed of vortex fluid, the vortex motions driven by a temperature gradient will generate a Josephson voltage, hence a substantial Nernst signal will be
detected. Wang et al. \cite{Wang2001,Wang2003,Wang2006} have carried out systematic measurements of Nernst
signal in HTSC in underdoped and overdoped samples, below and above $T_c$, in weak and strong fields, which
constitute a valuable source for understanding the carrier transport. Carefully documented data show that the
Nernst signal vary with $T$ and $B$ as a tilted-hill profile at $T<T_v$ and $H<H_{c2}$, where $T_v$ and
$H_{c2}$ are the onset temperature and field of Nernst signal, called upper critical temperature and field,
respectively. An important discovery in these measurements is the universal continuity of the signal below and above $T_c$. Several theoretical approaches have been developed to explain the Nernst effect in HTSC
(especially above $T_c$), but a quantitative description over a wide range of $T$ and $B$ is not yet
available.

Generally speaking, there are three microscopic components in HTSC, namely coherent (or condensed) Cooper
pairs, fluctuating Cooper pairs and quasiparticles. In a transport experiment, these components are driven to
form three macroscopic currents, i.e., vortex fluid, normal current of fluctuating pairs and quasiparticle
current. Consequently, three theoretical approaches can be identified, depending on the component to
emphasize, namely phase-fluctuations approach, Gaussian (i.e., amplitude)-fluctuations approach and
quasiparticle approach.

According to phase-fluctuations scenario, a spin-charge locking mechanism in pseudogap state leads to vortex
formation \cite{Anderson2006,Weng2006}. The large  positive Nernst signal above $T_c$ is interpreted as a
strong evidence for the existence of vortices in the pseudogap state \cite{Emery1995}. Based on the theory of
Berezinskii-Kosterlitz-Thouless (BKT) phase transition \cite{Berezinskii1972,Kosterlitz1972,Kosterlitz1973},
the phase-fluctuations scenario suggests that the superconductivity is terminated at $T_c$ by the loss of a
global phase coherence as a result of the activation of thermal vortices; nevertheless, local phase coherence
in the form of vortex motions is still present, driving a substantial Nernst signal. Following this idea,
$T_v$ (the onset temperature of Nernst signal) is also the onset temperature of local phase coherence and
local superfluid density. Simulation of phase fluctuations appropriately predicts a sharp temperature decay of Nernst signal above $T_c$ but fails to describe the behavior at low temperature \cite{Podolsky2007}. Note that the studies of nonlinear time-dependent Ginzburg-Landau (TDGL) equation \cite{Mukerjee2004,Tinh2009,Andersson2010,Chung2012} also predict the Nernst signal, but only signal below
$T_c$ in overdoped La$_{2-x}$Sr$_x$CuO$_4$ (LSCO) is reasonably described, while many data of other
samples in other range of $T$ and $B$ are yet to be compared.

Note that the original theory of vortex scenario was intensely challenged by others. Scenario of Gaussian
fluctuations is another candidate for discussing the origin of Nernst signal above $T_c$. It is analyzed by
Ussishkin et al. in low fields \cite{Ussiskin2002}, and the prediction $\alpha_{xy}\propto B(T/T_c-1)^{-1/2}$
compare favorably  with empirical data in overdoped LSCO. Later, this concept (Gaussian-fluctuations) is
extended to describe data at various $T$ and $B$ of NbSi and Eu-LSCO \cite{Serbyn2009,Michaeli2009,Chang2012}. However, the singular prediction near critical temperature and field due to the divergence of the coherence
length $\xi$ points to its limitation. A valid description with smooth transition between vortex fluid and  Gaussian-fluctuations at critical temperature and fields is still missing.

The quasiparticle approach is stimulated by recent discovery of density-wave order in various samples
\cite{Stock2006,Sebastian2008,Choini¨¨re2009,Fink2011,Atkinson2015},  and several models
\cite{Oganesyan2004,Zhang2008,Hackl2010} interpret Nernst effect in terms of transport  currents of
quasiparticles resulted from the Fermi surface reconstruction induced by density wave order.  The signal
predicted by Boltzmann equation can be made qualitatively similar to experimental data at  $T>T_c$, which,
however, is restricted to weak fields at high temperature, since the contribution of local superfluid is
neglected; in addition, quantitative comparison to experimental signals for all  regimes of $T$ and $B$ is
lacking. In summary, it seems that none of the above scenarios can yield a quantitative description of Nernst
signal over the entire range of temperature and magnetic field. Such quantitative description is important,
because it not only provides a unified picture of the transport properties in HTSC, but also helps to reliably extract physical parameters characterizing the vortex motions, and hence the pseudogap state. Finding such a
phenomenological description of the Nernst effect valid for all physical regimes is the goal of the present work.

In our view, the long standing controversy of Nernst effect in HTSC originates from an oversimplification of
the matter, i.e., insisting on finding a simple and single origin. In fact, HTSC provides a complicated
physical environment where there are interplays between several interacting components, in addition to
imperfect lattice and external fields. What one needs is a comprehensive theory synthesizing the three
currents, with quantitative description of $T$, $B$ and doping dependence of empirical data. The theory should address such  questions as what are dominant components in different regimes, and what are possible
transitions between them in the  phase diagram.

  It is widely recognized that vortex fluid is a vital component in multiple regimes of the $H-T$ phase diagram in HTSC \cite{Blatter1994}, and can be specified \cite{Nelson1988} as disentangled vortex liquid which exhibits a superconducting response along the direction of magnetic field and entangled vortex liquid
  which has the same symmetries as the normal phase. Anderson believed that the  Nernst signal in pseudogap phase is controlled by thermally excited, fluctuating, quantized vortex  tangles \cite{Anderson2007}, and went further to propose a concrete model of vortex dynamics  \cite{Anderson20062}, in which the transport entropy $S_{\phi}$ was chosen to satisfy the critical condition of BKT phase transition, and  the damping
 viscosity $\eta$ was determined from vortex-vortex collisions in a quantum- and thermal- fluctuations
 scenario. This model predicts the tilted-hill profile vs $B$, but a careful examination reveals a
 three-orders of magnitude difference between predicted and measured signal, once proper physical parameters are inserted in.

In order to resolve this controversy, we note that in the vortex dynamics view, both Nernst effect and flux-flow resistivity originate from the phase slippage induced by vortex currents
\cite{Anderson1966,Huebener2002}, and thus $\eta$ in these two cases should be similar. Using $\eta$ estimated from magnetoresistance experiment with a flux-flow resistivity model, we have revised Anderson's model, and then obtained very satisfactory description of both magnetoresistance and Nernst signal in six samples over a
wide range of $T$ and $B$. The success gives a strong support to the underlying picture of vortex-fluid
dynamics in pseudogap phase proposed by  Anderson  \cite{Anderson2006,Anderson20062,Anderson2007}.

Note that our quantitative vortex-fluid model is phenomenological, and describes an integrated picture of
phase and amplitude fluctuations, while some effect of quasiparticles (e.g. dissipation in the core) is also
taken into account. In previous studies, vortex motions are believed to originate only from phase fluctuations \cite{Podolsky2007}, which is indeed the case in underdoped samples. However, the successful description of
Nernst signal let us suggest that in overdoped samples, amplitude fluctuations generate (fluctuating) vortices (see more discussion in Sec. \ref{DisCon}), whose effect can also be correctly captured by the present model.
Therefore, our vortex fluid model offers an integrated picture of phase and amplitude fluctuations. Note also
that our picture does not completely remove the contribution of quasiparticle current; in fact, quasiparticles inside the vortex core move with the vortex and form a dissipative quasiparticle current, which is the origin of $B_0$ term below.

The paper is organized as follows. In Sec. \ref{FFR}, an integrated model is deduced to quantify the damping viscosity $\eta$ of vortex flow induced by impurity scattering of quasiparticles, vortex-vortex collisions and pinning. The validity of this model is quantitatively verified with magnetoresistance data of $\rm Bi_{2.1}Sr_{1.9}CaCu_2O_{8+\delta}$. In Sec. \ref{NerEff}, Anderson's model is revised  by a vortex damping
$\eta$ constructed in the former section. The discrepancy of three orders of magnitude between data and
Anderson's original model is pointed out and corrected.  The model yields predictions of the $T$ and $B$
dependence of Nernst signal, in quantitative agreement with data of six samples in Bi-2201 and Bi-2212. The
two-step procedure of parameter determination, {verified to be reliable for most parameters, is described in
the Appendix \ref{ParaDeter}, which enables to predict several key quantities such as local superfluid density and upper critical field and temperature.  Open questions about local superfluid density, the origin of
damping viscosity, and the effect of amplitude fluctuations are discussed in Sec. \ref{DisCon}.

Some acronyms are used to identify the cuprates, Bi-2201 for Bi$_{2}$Sr$_{2 - y}$La$_{y}$CuO$_{6}$, Bi-2212 for Bi$_{2}$Sr$_{2}$CaCu$_{2}$O$_{8+\delta}$, UD, OP, and OV stand for underdoped, optimally doped, and overdoped, respectively.

\section{flux-flow resistivity}\label{FFR}

\begin{figure}
  \centering
  \includegraphics[width=8cm]{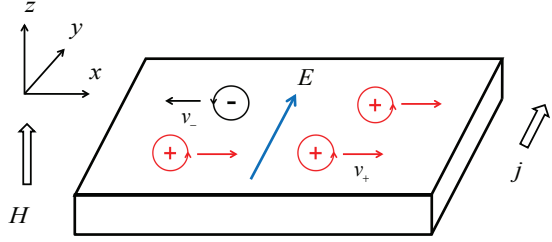}\\
  \caption{Schematic picture of the electric field $E$ induced by the vortex flow under a driven current $j$. There are two types of vortices and velocity $v_+=-v_-$, where $+$ and $-$ represent vortex and antivortex
  respectively.}
  \label{VortexFlowJ}
\end{figure}

It is known that the BKT phase transition which is found in superfluid helium films also occurs in quasi-2D
thin superconducting films and cuprates superconductors \cite{Jose2013}. This transition implies that the
vortex-antivortex pair bound at low temperature dissociates into free thermal vortices above a characteristic
temperature, which is chosen as $T_c$ in the present work. When magnetic field is applied in HTSC, there are
two types of vortices, namely magnetic and thermal vortices, whose density will be denoted as $n_B=B/\phi_0$
and $n_T$. Since the thermal vortices appear in pairs, the total number of vortices is $n_+=n_B+n_T/2$, and
that of antivortices (with a reversed  angular momentum) is $n_-=n_T/2$.

In the experiment of flux-flow resistivity as schematically described in Fig. \ref{VortexFlowJ}, a
transport current density \textbf{j} is applied $\parallel$ \textbf{\^{y}} and magnetic field
\textbf{H}$\parallel$ \textbf{\^{z}}. When the circular current and magnetic field of the vortex are changed
by \textbf{j}, a Lorentz force \textbf{f}$_L$=\textbf{j}$\times\phi_0/c$ is generated to form a vortex current perpendicular to \textbf{j}, where $c$ is the speed of light. The force \textbf{f}$_L$ must be balanced by a
damping force, namely
\begin{eqnarray}
{\bf j}\times \frac{{\bf\phi}_0}{c}=\eta {\bf v}_\phi,
\label{ForceBalance}
\end{eqnarray}
where $\eta$ is the damping viscosity per unit length and $v_\phi$ is the drifting velocity of vortices.
Vortex flow generates a phase slippage \cite{Anderson1966,Huebener2002},  in which the phase difference
between two arbitrary points A and B changes by $2\pi$ whenever a vortex crosses the link  AB. The phase
change leads to a Josephson voltage $2eV_J=\hbar\dot{\theta}$ between two sides of the sample, where
$\dot{\theta}=2\pi\dot{N_+}$, and $\dot{N_+}=n_+ Lv_{\phi}$ is the number of the vortices crossing the width
per unit time ($L$ is the width of the sample). Finally, the vortex flow generates a transverse electric field $E_+=n_+v_{\phi}\phi_0/c$. Meanwhile, the antivortex will transport in the opposite direction with
$v_-=-v_\phi$, which induces  an electric field $E_-=n_-v_\phi\phi_0/c$. The total electric field induced by
vortex and antivortex flow is thus $E=n_vv_\phi\phi_0/c$, where $n_v=n_B+n_T$ is the total density of
vortices. Substitute this relation and Eq. (\ref{ForceBalance}) into Ohm's law \textbf{E}=\textbf{j}$\rho$,
the flux-flow resistivity $\rho$ can be derived as
\begin{eqnarray}
\rho=\frac{n_v}{\eta}\frac{\phi_0^2}{c^2}.
\label{Resistivity-formula}
\end{eqnarray}
In Eq. (\ref{Resistivity-formula}), two key parameters are involved, namely the damping viscosity $\eta$ and
the vortex density $n_v$. In the following, we model $\eta$ first and then discus $n_v$.

\subsection{Model and formalism}

Generally speaking, damping viscosity $\eta$ involves several mechanisms, which are difficult to calculate
from the first principle. Phenomenological consideration seems to be the only promising way at this stage of
the knowledge. We assume that each single vortex is a circular supercurrent surrounding a normal core, which
is similar to the Bardeen and Stephen picture for local superconductor \cite{Bardeen1965}. Then, three damping mechanisms are considered, namely impurity (or defect) scattering of quasiparticles inside the vortex core,
vortex-vortex collisions, and an interaction between vortex and pinning center. Impurity scattering happens at the mean free path of quasiparticles which is the microscopic scale. The latter two works in two steps:
firstly, the deformation and transformation of the supercurrent of the vortices, thus at the scale of the
inter-vortex distance; second, a damping via collisions between carriers and lattice (impurity, phonon and pinning center). We assume that these three mechanisms can be decomposed so that $\eta$ can be expressed as
their linear superpositions,
\begin{eqnarray}
\eta=\eta_0+\eta_v+\eta_{pin},
\label{eta}
\end{eqnarray}
where $\eta_0$, $\eta_v$ and $\eta_{pin}$ are the contribution of impurity scattering, vortex-vortex collision and pinning, respectively. The crucial difference between HTSC and conventional SC is that $\eta$ for HTSC is
different in different phase space of $T$ and $B$. Specifically, in a conventional and clean superconductor,
vortex moves nearly independently so that the impurity scattering is dominant, while in a HTSC, vortex-vortex
collisions are more important in a wide range of parameter space due to much higher vortex density under small coherence length (see Sec. \ref{VVI}).

\subsubsection{Impurity scattering of quasiparticles}

HTSC is widely believed to be a doped Mott insulator \cite{Lee2006} in which the doping plays a crucial role in determining the transport property. Since the number of quasiparticles inside vortex core
(about 10 holes in OP Bi-2212) and the phonon density at low temperature are both small, the scattering
between phonon and  quasiparticles inside vortex core is negligible in a vortex fluid below $T_c$.
In this case, damping inside vortex core is dominated by impurity scattering. The impurity scattering is
approximately temperature-independent, thus $\eta_0$ should be a constant parameter of the material. Following the Bardeen-Stephen model \cite{Bardeen1965}, it can be expressed as $\eta_{0}=H_{c2}\phi_0/\rho_{res}c^2$
where $\rho_{res}$ is residual resistivity due to impurity scattering.

\subsubsection{Vortex-vortex collisions}\label{VVI}

First, we discuss the regime below $T_c$ where thermal vortices are absent. The independent vortex assumption
is valid in low fields but inappropriate in high fields in HTSC. For cuprates superconductors, $H_{c2}$ are in the range of $10\sim $100 T. When the sample is tested in high fields (say $B$=10 T), the distance between
vortex cores $l_B=\sqrt{\Phi_0/B}$ becomes of the order of 100 \AA, which is much smaller than the penetration depth $\lambda\sim1000$ \AA. For a single vortex, the circular supercurrent {\textbf{j}}$_s$ survives outside
the core at the length scale of $\lambda$ so that there is a huge overlap of the supercurrent of adjacent
vortices. This may induce collisions to enhance the damping. In this process, vortices collide with each
other, which leads to a transformation from circular supercurrents to fluctuating Cooper pairs and
quasiparticles, and then the motions of the two products are damped  by collisions with lattice (impurity and phonon).

When only two-body collisions are considered, the collision rate is linearly proportional to the vortex
density, thus a linear $B$ dependence of $\eta_v$ can be expected,
\begin{eqnarray}
\eta_{v}&=&\frac{\phi_0}{c^2}\frac{B}{\rho_0},
\label{etav}
\end{eqnarray}
where $\rho_0$ is a characteristic resistivity describing the damping strength due to a single process of
two-body collision. Since vortex-vortex collisions dominate in most regime of HTSC phase diagram, $\rho_0$ is
good parameter to quantify damping viscosity $\eta$. Express $\eta_0$ in terms of $\rho_0$, we obtain
\begin{eqnarray}
\eta_0=\frac{\phi_0}{c^2}\frac{B_0}{\rho_0},
\label{eta0}
\end{eqnarray}
where $B_0=H_{c2}\rho_{0}/\rho_{res}$ is the effective fields to describe the damping strength of impurity
scattering relative to vortex-vortex collision.

\subsubsection{Pinning effect}

Pinning effect introduces modification to the transport of a vortex fluid \cite{Blatter1994}. If vortex fluid
passes by a pinning center, the vortex interacts with pinning center, thus the vortex core is slowed down, and the current density in the core is not equal to the applied transport current density. At low temperature,
Blatter et al. developed a phenomenological theory based on an assumption of weak collective pinning
\cite{Blatter1994}. We follow their main picture and deduce a quantitative model of pinning.

In a pinned vortex fluid, the Lorentz force is balanced by the pinning and damping force,
\begin{eqnarray}
(\eta_0+\eta_v) {\bf v}_{\phi}+{\bf f}_{pin}={\bf f}_L.
\label{dampingforce}
\end{eqnarray}
The pinning force can be expressed in an equivalent form $f_{pin}=\eta_{pin} v_{\phi}$, where $\eta_{pin}$  is the effective viscosity.  Since pinning and vortex-vortex collisions both works at inter-vortex distance,
their joint effect can be modeled by a multiplicative factor $\Gamma$, such that $\eta_v+\eta_{pin}=\Gamma\eta_v.$ The value of $\Gamma$ must satisfy two limiting conditions. First, at high
temperature and strong fields where the pinning effect vanishes, $\Gamma$ is equal to 1.  At low temperature
and weak field regime,  it is the thermal assisted flux flow (TAFF) described by the classic exponential form
(Arrhenius law) $\rho\approx \rho_0\exp[-U_{pl}/(k_BT)]$, where $U_{pl}$ is the plastic deformation energy
\cite{Kes1989, Blatter1994}. A simple choice is thus $\Gamma=\exp(U_{pl}/k_BT)$, which yields the following
expression:
\begin{eqnarray}
\eta_{pin}=\frac{\phi_0}{c^2}\frac{B}{\rho_0}[\exp(U_{pl}/k_BT)-1].
\label{etap}
\end{eqnarray}
In summary, the resistivity of pinned vortex fluid below $T_c$ is
\begin{eqnarray}
\rho= \rho_0 \frac{B}{B_0+B \exp(U_{pl}/k_BT) }.
\label{ResistivityLowT}
\end{eqnarray}

The $B$ and $T$ dependence of $U_{pl}$ may vary with materials, and even with samples \cite{Blatter1994,Tinkham1988}. So, it should be parameterized. In this work, we will follow Geshkenbein et
al.[43] to make an approximation in bismuth compound. In the collective pinning theory\cite{Blatter1994,Larkin1973}, the vortex is pinned by the collective action of many weak pointlike
pinning  center via deformation of the vortex structure.  Geshkenbein et al. \cite{Geshkenbein1989} proposed a form of $U_{pl}$ in terms of the energy involve deformations of the vortex lines on a scale $l_0$:
 $U_{pl}\propto\gamma^{-1}\varepsilon_0l_0$, where $\gamma$ is an anisotropic parameter, $\varepsilon_0$ is
 the characteristic energy of unit length of a single vortex, which is proportional to the 2D superfluid density $n_s$.  Furthermore,  taking  the usual linear dependence of the superfluid density, $n_s= n_{s0}(1-T/T_v)$, and with  $l_0=l_B$ (the inter-vortices distance), we obtain
 $n_sl_0=n_{s0}(1-T/T_v)\sqrt{\phi_0/B}$, where $n_{s0}$ is a characteristic superfluid density and $T_v$ is
 the onset temperature of local superfluid density.  We introduce a parameter $B_p$ to express pinning
 strength. Then, the plastic barrier can be expressed as
\begin{eqnarray}
U_{pl}=k_B(T_v-T)\sqrt{B_p/B}.
\label{Upl}
\end{eqnarray}
In experiments of BSCCO, this typical magnetic field \cite{Ferrari1990} and temperature \cite{Kucera1992}
dependence are both discovered. $B_p$ is an intrinsic material parameter, which can be probed by the melting
field $H_m$ as we derive now. Based on a dynamic approach \cite{Blatter1994}, the melting field $H_m$ obtained from the Lindemann criterion \cite{Lindemann1910} can be expressed via the plasticity energy barrier $U_{pl}$
at $B=H_m$, $k_BT=2c_L^2U_{pl}$, {where $c_L\sim0.1-0.4$ is the Lindemann number. Then, $H_m$ measured from
data can be used to determine $B_p$, by solving it from Eq. (\ref{Upl}) at $B=H_m$, namely
\begin{eqnarray}
B_p={H_m}/(4c_L^{4} t^{2}).
\label{Bp}
\end{eqnarray}
where $t=T_v/T-1$.}

\subsubsection{Thermal vortices} \label{ThermalVortices}
Above $T_c$, the total density of thermal vortices, $n_{T}=2n_-$, can be determined by vortex correlation
length $\xi_+$ as $n_{T}=\xi_+^{-2}$, where $\xi_+$ represents the characteristic scale beyond which thermal
vortices begin to unbind \cite{Kadin1983}.  Using a renormalization group technique, Kosterlitz found that as
$T\rightarrow T_c$,  the correlation length diverges asymptotically like \cite{Kosterlitz1974}
\begin{eqnarray}
\xi_+\sim\exp[b{(T/T_c-1)^{-1/2}}].
\label{xi}
\end{eqnarray}
In literature, this temperature dependence was verified in the critical regime near $T_c$ and $b$ is found to
be non-universal \cite{Hebard1983,Martin1989,Ying1990}. In fact, the Nernst region above $T_c$ is the lower
part of pseudogap phase (UD regime) and strange metal phase (OP and OD regime), where superconducting
fluctuations dominate the physical properties. Thus, we speculate that the superconducting fluctuations make
thermal vortices undergo a critical behavior. In other words, Eq. (\ref{xi}) of $\xi_+$ should be applicable
at $T_c<T<T_v$. Therefore, we can define the following effective fields for modeling effects of thermal
vortices:
\begin{eqnarray}
B_{T}=\Phi_0/\xi_+^2=B_{l}e^{-2b|T/T_c-1|^{-1/2}},
\label{BT}
\end{eqnarray}
where $B_{l}$ represents the high-temperature limit of $B_{T}$. At this limit, cores of adjacent  thermal
vortices overlap with each other, i.e., correlation length of thermal vortices  is equal to the coherence
length $\xi_+=\xi$. In a special situation when $b\ll 1$, $B_{T}$ saturates quickly to $B_{l}$ at $T>T_c$, so
that the critical state is immediately reached near $T_c$, then $B_{l}\approx H_{c2}'$, where $H_{c2}'$ is the upper critical field at $T=T_c$. In this paper, we use this approximation to achieve the comparison between
theoretical predictions and experimental data.

Taking the contribution of thermal vortices into consideration above $T_c$, the total damping viscosity can be
written as
\begin{eqnarray}
\eta= \frac{\phi_0}{\rho_0c^2}\{B_0+(B+B_{T})\exp[t\sqrt {{B_p}/(B + {B_{T}})}]\}.
\label{eta}
\end{eqnarray}
 Here, we have replaced $B$ in Eq. (\ref{Upl}) by $B+B_{T}$. The integrated model Eq. (\ref{eta}) of $\eta$ enables us to obtain an analytic model of flux-flow resistivity:
\begin{eqnarray}
\rho= \rho_0 \frac{B+B_{T}}{B_0+(B+B_{T})\exp[t\sqrt {{B_p}/(B + {B_{T}})}]}.
\label{Resistivity}
\end{eqnarray}

\subsection{Comparison to experiment}\label{rhocompSec}

\begin{table*}[tph]
\caption{Material parameters in Eq. (\ref{Resistivity}) determined in two steps, rough estimate (RE) and fine
tuning (FT), described in Appendix \ref{ParaDeter}, for $\rm Bi_{2.1}Sr_{1.9}CaCu_2O_{8+\delta}$.\cite{Usui2014}
The fine tuning reduces the root mean square error(RMSE) from
 $\approx10\mu\Omega$cm (about 10\% of $\rho_0$) to $\approx5\mu\Omega$cm (about 5\% of $\rho_0$ or 2 times
 smaller); the latter is the experimental uncertainty.}
\begin{center}
\renewcommand\arraystretch{1.5}
\begin{tabularx}{\textwidth}{ >{\setlength{\hsize}{.7\hsize}}X >{\setlength{\hsize}{.9\hsize}}X
>{\setlength{\hsize}{1.2\hsize}}X >{\setlength{\hsize}{0.9\hsize}}X >{\setlength{\hsize}{0.9\hsize}}X >{\setlength{\hsize}{0.9\hsize}}X >{\setlength{\hsize}{0.9\hsize}}X >{\setlength{\hsize}{1.1\hsize}}X
>{\setlength{\hsize}{1.5\hsize}}X}
\hline
\hline
Step & $T_c$ (K) &  ${\rho _0}$ (m$\Omega$cm) & $T_v$ (K) &$B_0$ (T) &$B_p$ (T) & $B_{l}$ (T) &b       &RMSE ($\mu\Omega$cm)\\
\hline
RE & 87       &  0.116                        & 112      & 4      &   30.8  & 68.1       & 0.42  & 10\\
\hline
FT & 87       &  0.99$\pm$0.11   & 104$\pm$3  & 0.4$\pm$0.4  & 89$\pm$34  & 68.1  & 0.56$\pm$0.11  & 5\\
\hline
\hline
\end{tabularx}
\end{center}
\label{ParaRes}
\end{table*}

Recently, in-plane magnetoresistance of oxygen-controlled Bi-2212 single crystals was measured by Usui et
al. \cite{Usui2014}. In OP ($p=0.16$) sample of $\rm Bi_{2.1}Sr_{1.9}CaCu_2O_{8+\delta}$, $\rho$ is measured
in a wide range of temperature (20-120 K) and fields (0-17.5 T). As is shown in Fig. \ref{rhocomp}, $\rho$
first shows a slow increase regime at low temperature, which is a TAFF regime, and then the curves converge at a characteristic temperature (i.e., $T_v$). The transition between the two regimes varies with $H$.

In order to verify the integrated model Eq. (\ref{Resistivity}),  a reliable procedure must be developed to
determine (seven) physical parameters.  A two-step procedure is carried out, with a rough estimate first and
then a fine tuning (see Appendix \ref{rhopara} for a detailed presentation). This determination is not a pure
mathematical fitting, but by a procedure with careful and thorough physical analysis, which ensure their
reliability.} The main idea can be summarized as follows. First, a rough estimate (RE) is conducted using
asymptotical analysis at large and small limits of $T$ and $B$ where only one or two parameters are relevant { and easily identifiable on the curves according to their physical meaning.  Then, a fine tuning (FT) process
searches for optimal parameters around the RE values, which minimizes the errors between a set of predictions
and data. At the FT step, precise parameter values and its uncertainties are obtained, see Table
\ref{ParaRes}.  The errors are generally very small so that very good agreement between experimental data
(symbols) and the predictions (solid line) is achieved, as shown in Fig. \ref{rhocomp}, for all five magnetic
field values. The agreement is specially good at $H$=0, i.e., about the thermal vortices. These results
validate our model of $\rho$.

Let's discuss further in detail the reliability of each parameter determination. First, the upper critical
temperature $T_v$ is the onset temperature of short-range coherence and thermal vortices, when decreasing the
temperature. In the vortex-fluid model, it indicates the temperature, beyond which the magnetoresistance
collapses and Nernst signal vanishes. Using the data of Usui et al., $T_v$=112 K is obtained in RE and
corrected to $104$ K in FT, and the two values are very close. Applying this procedure to other data of OP
Bi-2212, we find $T_v$ is nearly constant for all samples. For example, according to an examination of the
magnetoresistance data and Nernst signal measured by Ri et al. \cite{Ri1994}, we find  $T_v$ in the range of
$[100,110]$ K for both. In the next section, we will show that $T_v$ can be derived from the Nernst signal of
Wang et al, which is $107\pm1.6$ K. All these results indicate that $T_v$ is the intrinsic parameter in HTSC
samples, and can be well defined. $T_v$ provides  a definition for $\rho_0$ as well, which is the high-
temperature and field limit of $\rho$, i.e., the usual normal-state resistivity. It is natural to define
$\rho_0$ at $T_v$: $\rho_0\approx\rho(T_v)$ as a rough estimate. Then, the robustness of $T_v$ also implies
the robustness of $\rho_0$.

$B_0$ is the contribution from  the quasiparticles scattering(inside vortex core) by defects. In the RE, we
propose to determine $B_0$ at $T=T_c$ where the vortex-vortex collisions and pinning effect have minimum
effects. Obviously, the reliability of the procedure depends on whether sufficient data of $\rho$ vs $H$¡¯ near the minimum location are available. Presently, only data at $H=$ 0.5 and 1 T are used to perform the RE, and
the result is $B_0=4$ T. Then, $B_0$ is found to be 0.4$\pm$0.4 T in the FT. This difference is rather
significant, i.e., the  reliability of the determination of $B_0$ in this sample is not good, which is
attributed to two factors. First, the data are scarce. Second, $B_0$ is too small. A reliable determination is possible when sufficient data of $\rho$ vs $H$¡¯ are available in the sample where the scattering between
individual vortex and defects is strong.

$B_p$ is the parameter to determine pinning strength, and thus can be independently determined with data in
TAFF regime. In the RE procedure, $B_p=30.8$ T is obtained, while $B_p=89\pm34$ T is obtained after applying
the FT method. This big difference is attributed to both the measurement error and the uncertainty of the
model. Firstly, $\rho$ is exponentially small in the TAFF regime,  so its measurement has a low signal to
noise ratio. This significantly affects the FT procedure, which requires a significant error sensitivity to
work. Secondly, $T_v$ (112 K) in RE is overestimated compared to $T_v$ (104 K) in FT, which, according to Eq.
(\ref{Upl}), yields an underestimate of $B_p$. Thirdly, below $T_c$, $\rho$ vs $T$ only depends on
$U_{pl}/k_BT\propto(T_v/T-1)$. As shown in Fig. \ref{rhocomp}, it is qualitatively right, but with
quantitative deviation, especially at low temperature (the TAFF regime). This can be improved by a revised
Geshkenbein et al.'s model, e.g., $U_{pl}/k_BT\propto(T_v/T-1)^{0.8}$. The non-integer index 0.8 quantifies
the crossover of states in TAFF regime, which will be discussed elsewhere.

Furthermore, the deviations between data and prediction at low field ($H=0.5$ T) below $T_c$ reveal another
defect of Geshkenbein et al.'s model, namely an overestimate of vortex deformation length $l_0$ by the
inter-vortex distance $l_B$ in Eq. (\ref{Upl}). At $H=0.5$ T, $l_B\approx640$ \AA\ while an
accurate fitting requires $l_0\approx370$ \AA. This difference is in fact physical, because of the variation
of the pinning  mechanism from weak to strong fields. In  weak fields, the pinning  effect is generated by the
interaction between isolate vortex and pinning cites. Since the vortex line cannot resolve lengths smaller
than coherence length $\xi$ ($22$ \AA\  in OP Bi-2212), the vortex deformation happens above the scale of
$\xi$\cite{Geshkenbein1989}. Thus, Eq. (\ref{Upl}) is invalid at weak-field limit. While at strong-field
limit, the strong vortex-vortex interaction forms an elastic vortex lattice which is a starting point of the
vortex fluid. In this case, pinning sites deform lattice and the deformation length is the lattice constant
$l_B$. In other words, when magnetic field increases from weak to strong, the vortex deformation length $l_0$
increases from $\xi$ to $l_B$. This can by expressed by a bridge function
$\l_0=\xi_0(1+B/B')^{-1}+l_B(1+B'/B)^{-1}$, where $B'=0.39$ T is chosen to satisfy $l_0=370$ \AA. Applying
the revised $l_0$ in the model, the fitting at $H=0.5$ T become very precise, which will also be reported
elsewhere.

In the determination of $B_l$, we assume that the high-$T$ limit of thermal vortex density is as dense as
magnetic vortices at $H_{c2}'$. To validate this assumption, one feasible idea is to analyze the thermal
vortices density in the numerical simulation. In the  simulation using the TDGL \cite{Chung2012},  Chung et
al. found that the density of thermal vortices is affected by magnetic field, and the vortex-proliferation
temperature decreases when field increases. Physically, $H$ tends to suppress the pair dissolving energy, thus
favor the dissociation of vortex-antivortex pair, similar to the electrostatic induction in which external
electric field induces a separation of negative and positive charge \cite{Minnhagen1987}, thus decreases
$T_c$. Maybe, it more or less relates to the deviations in high fields such as $H=17.5$ T and 9T near $T$=80 K in Fig. \ref{rhocomp}. The BKT physics at finite magnetic field has been theoretically investigated in recent
years \cite{Oganesyan2006,Benfatto2007,Wachtel2014}, and the nonlinear field dependence of the total density
of free vortices has been found. However, one can find with Debye-H\"{u}ckel approximation that the nonlinear
field dependence is only remarkable when $n_T\sim n_B$ \cite{Wachtel2014}, and is thus neglected as a higher
order effect in the present work.

$b$ is the unique parameter to determine the $T$ dependence of thermal vortex activation. And, the value $b=0.42$ is obtained in the RE, and 0.56$\pm$0.11 in the FT; both are close. The validity
of this result can be further confirmed by applying the classic procedure \cite{Martin1989}, where a synthetic model of Halperin-Nelson model \cite{Halperin1979} and Bardeen-Stephen model. In order to compared with our
model, the constants in the original expression \cite{Martin1989} at $T_c$ are combined, thus
\begin{eqnarray}
\rho  \propto {\rho _0}\exp \{ - 2b(T/T_c - 1)^{ - 1/2}\} .
\label{rhoHN}
\end{eqnarray}
Follow the usual procedure, we fit the data of $\rho$ in zero field for $\rm Bi_{2.1}Sr_{1.9}CaCu_2O_{8+\delta}$ \cite{Usui2014} and obtain $b=0.55$. The agreement between the two results is not surprise. In zero field, our model is
\begin{eqnarray}
\rho  = {\rho_0}\frac{B_T}{B_0+ B_T\exp (t\sqrt {B_p/B_T} )}.
\label{rhozero}
\end{eqnarray}
This model is well verified by the data in zero field in Fig. \ref{rhocomp}. The only difference between Eq.
(\ref{rhoHN}) and Eq. (\ref{rhozero}) is that, ${B_T}\exp(t\sqrt{B_p/B_T})$ in Eq. (\ref{rhozero}) is replaced by a constant in Eq. (\ref{rhoHN}). As we discussed in Appendix \ref{rhopara}, ${B_T}\exp(t\sqrt{B_p/B_T})$ is almost a constant near its minimum, thus the $T$ dependence of $\rho$ is mainly determined by Eq.
(\ref{rhoHN}). But, when $T$ increases away from the minimum, ${B_T}\exp(t\sqrt{B_p/B_T})$ increases. The
neglect of this variation is the reason for the slight overestimate of $b$ from Eq. (\ref{rhoHN}). In a word,
taking the contribution of collisions between thermal vortices in the damping viscosity enables us to determine $b$ more accurately by Eq. (\ref{rhozero}).

Note that Martin et al. \cite{Martin1989} determined $b=0.183$ using Eq. (\ref{rhoHN}) in a 2D (of 2 $\mu$m
thick) and nearly optimal doped sample, where $b$ is much smaller than our results in Usui et al.'s sample.
Since smaller $b$ represents a stronger activation of thermal vortices, this enables us to speculate that the
low dimension effect (i.e., strong fluctuations) may strengthen the thermal vortex activation. In order
to validate this idea, it is intriguing to arrange a set of samples with different thicknesses, and our
formula will be useful to quantify the thickness dependence, through the variation of $b$.

In summary, the magnetoresistance measured by Usui \cite{Usui2014} is well described by the vortex-fluid model, Eq. (\ref{Resistivity}). Since $T_v$ and $\rho_0$ are important characteristics in experimental curves,
their determination is robust and reliable. The comparison with data shows the variation above $T_c$ is
accurately described, as well as the Kosterlitz coefficient $b$; but the quantitative deviation is present for the description of pinning  effect in the TAFF regime below $T_c$, which results in an ill-determination
of $B_p$. Besides, $B_0$ is too small in this sample to be reliably determined.

\begin{figure}
  \centering
  \includegraphics[width=7cm]{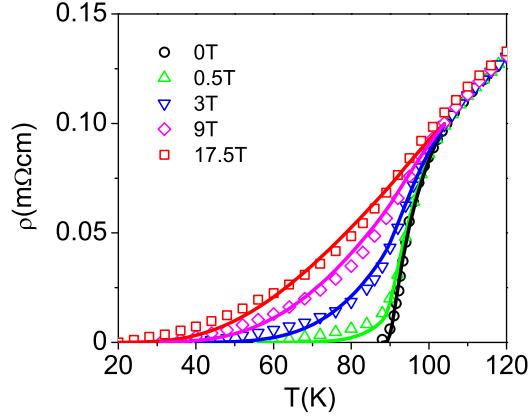}\\
  \caption{Comparison between predictions (solid line) and experimental data (symbols) \cite{Usui2014} using the fine tuned parameters  given in Table \ref{ParaRes} at $H$=0, 0.5, 3, 9, 17.5 T.  Only the data at $T\leq104$ K are fitted.}
  \label{rhocomp}
\end{figure}

\section{Nernst Effect}\label{NerEff}

\begin{figure}
  \centering
  \includegraphics[width=8cm]{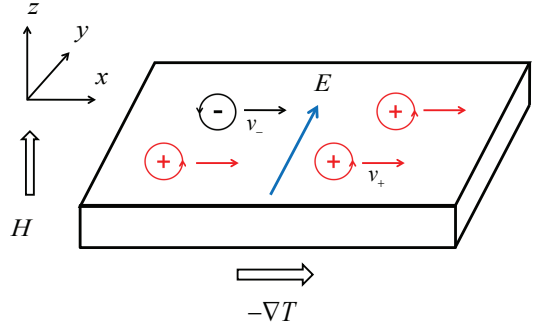}\\
  \caption{Schematic diagram of Nernst effect. Vortices and antivortices are driven by temperature gradient
  $\nabla T$ to form a flux flow, and then results in a Josephson voltage perpendicular to $\nabla T$.}
  \label{VortexFlowT}
\end{figure}

According to the two-fluid model of superconductivity, Nernst signal can be decomposed into quasiparticle
signal $e_N^{n}$ and signal $e_N^{SC}$ of superconducting order parameter $\psi$. In this paper, we quantify
the vortex fluid contribution, which  captures contributions of phase and amplitude fluctuations (see
introduction in Sec. \ref{DisCon}).

In Fig. \ref{VortexFlowT}, a temperature gradient $-\nabla T$ is applied $\parallel$ \textbf{\^{x}} and
magnetic field \textbf{H}$\parallel$\textbf{\^{z}}. As vortex contains higher entropy(land heat) than
surrounding fluid, a thermal force \textbf{F}$_{th}$ generated by $-\nabla T$ acts on vortices. This results
in a heat transport with an energy $U_\phi=TS_\phi$, where $S_\phi$ is the transport entropy per unit length
of a vortex \cite{Huebener1969,Sergeev2010}. When a vortex is transported across a distance $\Delta x$, the
heat change $\Delta U_{\phi}=\Delta T S_\phi$ equals the work done by the thermal force -$F_{th}\Delta x$. So, we obtain
\begin{eqnarray}
{\bf F}_{th}=S_{\phi}(- \nabla T),
\label{Fth}
\end{eqnarray}
Thermal force should be balanced by damping force due to energy dissipation, which, when vortices move in uniform velocity \textbf{v}$_\phi$, is
\begin{eqnarray}
S_{\phi}(- \nabla T)= \eta {\bf v}_\phi.
\label{FBalanceN}
\end{eqnarray}
Thermal diffusion and the transport velocity is the same for vortices and anti-vortices as they have the same
thermodynamic properties (with only different angular momentum), as illustrated in Fig. \ref{VortexFlowT}. Due to the reversal of flux, the phase slippage of thermal vortices  and anti-vortices  act in opposite direction
and cancel with each other. Then, the total induced electric field is $E=n_Bv_\phi\phi_0/c$, indicating that
only the magnetic vortices contributes to Nernst signal. This yields $E =-B S_{\phi}\nabla T/\eta c$. Define
Nernst signal to be $e_N=E/(-\nabla T)$, then
\begin{eqnarray}
e_N= \frac{B}{c} \frac{S_{\phi}}{\eta}.
\label{NernsteN}
\end{eqnarray}

\subsection{Model and formalism} \label{eNMod}

$S_\phi$ and $\eta$ are the two key quantities to describe Nernst effect. The origin of the former still
presents some controversies \cite{Sergeev2010}. Nevertheless, the universal tilted-hill profile of Nernst
signal let us believe that the physics of $S_\phi$ must be simple. We now develop an order-disorder balance
argument for its derivation. In the superconducting state, the ordered motions (e.g., the supercurrents)
obviously dominate over disordered motions (e.g., the quasiparticle currents), while it is the opposite in the normal state. Then, we assume that the vortex-fluid state is an intermediate state where the ordered and
disordered motions achieve a balance. In other words, the key assumption in the present work is that, in the
vortex fluid, the energy of ordered motions (e.g. $E_k$, the kinetic energy of supercurrents around the core)
equals the energy of disordered motions (e.g. $U_{\phi}= TS_{\phi}$, encompassing the energy of random motions of the vortex core and of quasiparticles inside the core). Thus, our balance argument yields $TS_\phi={E_k}$, or
\begin{eqnarray}
S_\phi = {E_k}/T.
\label{Sphi}
\end{eqnarray}
Note that we neglect the ordered motions inside the vortex core, which is consistent to the cheap vortices
scenario proposed by Lee et al. \cite{Lee2006}. Eq. (\ref{Sphi}) is also consistent with Anderson's proposal
of expressing $S_\phi$ in terms of the critical condition of BKT phase transition near $T_c$
\cite{Anderson20062}. However, the balance argument is more general than the critical condition of BKT phase
transition, although Anderson has guessed it in his footnote \cite{Anderson20062}. Our proposal is supported
by the comparison of the model's prediction to experimental data (see below), but further more direct
verification would be intriguing.

Anderson claimed that magnetic vortices do not experience Kosterlitz-Thouless screening, and so the velocity
field retains the entire $1/r$ structure until reaching the magnetic length $l_B$. This is similar to the
Gauss theorem for charge. Therefore, the kinetic energy can be obtained by an integration from vortex core
size $\xi$ to $l=l_B/\sqrt{2\pi}$ ($\sqrt{2\pi}$ is the structure factor),
\begin{eqnarray}
E_k= \int_{\xi}^{l} {{n_s^*}\frac{{{m^*}}}{2}} {v_{sc}^2}{d^2}r,
\label{Ek}
\end{eqnarray}
where $n_s^*=n_s/2$ is the local density of Cooper pairs, $m^*=2m_c$ ($m_c$ is the effective mass of the hole
carrier), $v_{sc}=\hbar/(m^*r)$ is the superfluid velocity around the core. For simplicity, $m_c$ can be
assumed as electron mass $m_e$, thus the transport entropy per unit length is
\begin{eqnarray}
S_\phi= {C_1\over c_0} {n_s\over T}\ln{\frac{H_{c2}}{B}},
\label{Sphi2}
\end{eqnarray}
where $C_1={\pi\hbar ^2/{8m_e}}=4.76\times10^{-28}$ cm$^4$gs$^{-2}$, $c_0$ is the c-axis lattice constant.

Furthermore, same damping mechanisms of vortex are responsible for Nernst effect and flux flow, thus the analytic model
of $\eta$ in Eq. (\ref{eta}) is also valid for Nernst signal. Combining Eq. (\ref{eta}) and Eq. (\ref{NernsteN}), the prediction
of $e_N$ is:
\begin{eqnarray}
e_N = {e_T}\frac{{B\ln [{H_{c2}}/B]}}{{{B_0} + (B + {B_{T}})\exp [t\sqrt {{B_p}/(B + {B_{T}})} ]}},
\label{eN}
\end{eqnarray}
where
\begin{eqnarray}
e_T={C_1 c\rho_0\over\phi_0 c_0}{n_s\over T},
\label{eT}
\end{eqnarray}
is a $T$-dependent characteristic signal. Note that, since we take the same $\eta$ as in magneticresistance
experiments, an estimate of the thermal electrical coefficient $\alpha_{xy}=e_N/\rho$ is given by
\begin{eqnarray}
\alpha_{xy}= \alpha_T\frac{B}{{B + {B_{T}}}}\ln \frac{{{H_{c2}}}}{B}.
\label{alphaxy}
\end{eqnarray}
where $\alpha_T=e_T/\rho_0=(C_1 c/\phi_0 c_0)(n_s/T)$ is a $T$-dependent characteristic thermal electrical
coefficient. This formula offers a tool to test model of $S_{\phi}$, a topic which has attracted much
attentions in the literature \cite{Capan2002}.

\subsection{Magnitude correction to Anderson's model}\label{Correction}

Our model is a revised Anderson's model, as explained now. In his description of the damping mechanism,
Anderson assumed a random vortex fluid generated by quantum fluctuations. Like Brownian particle, pancake
vortices satisfy also the well know Einstein relation between the damping viscosity $\eta^p$ and diffusion
coefficient $D$: $D\eta^p=k_BT$ where $ \eta^p= c_0\eta$ is the total damping of one pancake vortex. Based on
the quantum fluctuations, the dimensional estimate yields $D= v_Dl_v=\hbar /m_v$, where $m_v$ is effective
mass the pancake vortex. Near $T_c$, $k_BT$ is substituted by kinetic energy of stochastic motion $v_D= \hbar^2/(m_vl_v^2)$, which yields
\begin{eqnarray}
\eta=\hbar n_v/c_0,
\label{etaAnderson}
\end{eqnarray}
Substituting this into Eq. (\ref{NernsteN}), Anderson obtained
\begin{eqnarray}
e_N=\frac{\pi\hbar}{8m_ec}\frac{n_s}{n_v}\frac{B}{T}\ln{\frac{H_{c2}}{B}}.
\label{eNAnderson}
\end{eqnarray}
Above $T_c$, Anderson assumed that thermal vortices dominate in $n_v$ so that $n_v$ is nearly $B$-independent.
Therefore this formula predicts a hill profile with large slopes in low fields and linear decrease to zero at
$B\approx H_{c2}$. These characteristics agree qualitatively well with the general shape of $e_N$ vs $B$.

However, a distinct magnitude problem in Anderson¡¯s model can be seen according to a rough estimate.
At $T_c$, thermal vortices vanish and $n_v=B/\Phi_0$, Then, Anderson¡¯s model predicts $e_N\approx2$ mVK$^{-1}$ for OP Bi-2201 and $e_N\approx1$ mVK$^{-1}$ for OP Bi-2212; these predictions are three orders of magnitude
higher than the experimental measured values, 5 $\mu$V{K$^{ - 1}$} and 3 $\mu$V{K$^{ - 1}$} \cite{Wang2003,Wang2006}. This discrepancy is in fact due to underestimated  damping viscosity $\eta$.
According to Anderson's argument of quantum fluctuations, $\eta=\hbar
n_v/c_0\approx2\times10^{-10}$ gs$^{-1}$cm$^{-1}$, but the magneticresistance experiment yields $\rho\approx
0.1$ m$\Omega$cm, from which one gets $\eta=2\times10^{-8}$ gs$^{-1}$cm$^{-1}$ for $B=1$ T with Eq. (\ref{Resistivity-formula}), which is two orders of magnitude larger than Anderson's estimate. This difference explains the most important part of the discrepancy. The additional one order of magnitude difference comes
from the over-estimation of $n_s$ and other numerical factors. The above reasoning let us think that the
diffusion is not dominated by quantum fluctuations, but by a complex interplay involving magnetic and thermal
vortices.

There are also two minor problems in Anderson's model. First, the scattering between vortex and lattice was
neglected in $\eta$, so a divergence is present in low fields at and below $T_c$ as $e_N\propto\ln(H_{c2}/B)$.
Second, $e_N$ varies monotonically with $T$, so that it cannot describe the non-monotonic
behavior of $e_N$ from high to low temperature. In conclusion, Anderson's model needs to be revised in order
to form a realistic model of viscous vortex fluid. In this paper, substituting the damping viscosity model Eq. (\ref{eta}) developed for magneticresistance, we propose a revised model Eq. (\ref{eN}) for describing Nernst signal, which not only predicts the right magnitude, but also displays sound behavior over a wide range of
temperature and field strength.

In order to predict the magnitude of signal, $n_s$ should be specified. In Sec. \ref{ThermalVortices}, we have mentioned that the local superfluid density $n_s$ vanishes at $T_v$. Here, we make a further assumption that
$n_s$ decays linearly from $n_{s0}$ at zero temperature to zero at $T_v$, thus, $n_s={n_{s0}}(1 - T/{T_v})$.
Then, $e_T$ in Eq. (\ref{eT}) can be expressed by
$e_T=e_c (T_v/T-1)$, where
\begin{eqnarray}
e_c={C_1 c\rho_0\over\phi_0 c_0}{n_{s0}\over T_v}.
\label{ec}
\end{eqnarray}
$e_c$ is a sample-dependent characteristic signal, which can be used to estimate the magnitude. In the
magnitude estimate with Eq. (\ref{ec}), $n_{s0}$ can be chosen as the hole density $n_h$, and $T_v$ can be
taken as $T_c$. Besides, the experimental data indicate $\rho_0\approx 0.1$ m$\Omega$cm \cite{Usui2014}. For OP Bi-2201 \cite{Zhou1999}, ${n_{h}}=5.56\times {10^{13}}$ cm$^{-2}$, ${c_0}=24.6$ \AA, ${T_c}=28$ K, then ${e_c}\approx19$ $\mu$VK$^{- 1}$. For OP Bi-2212, ${n_{h}}=10.9\times {10^{13}}$ {cm$^{-2}$}, ${c_0}=30.9$
\AA, $T_c=90$ K, then ${e_c}\approx9$ $\mu $V{K$^{-1}$}. These values are very close to the experimental
values of 5 $\mu$V{K$^{-1}$} and 3 $\mu$V{K$^{-1}$} \cite{Wang2003,Wang2006}. This agreement let us conclude
that not only Anderson's model for the entropy, but also our model for the damping viscosity chosen from
flux-flow resistivity are validated.

\subsection{Comparison to experiment}\label{eNcompSec}

\begin{table*}[tph]
\caption{Parameters determined from Nernst signal of six samples in Bi-2201 \cite{Wang2006} and
Bi-2212 \cite{Wang2003,Ong2008}. The italic number at the second line represents the roughly estimated value.
$T_v$, $B_0$, $B_p$, $B_{l}$ and $b$ are determined from Eq. (\ref{eN}) with error bar estimated at root mean
square error(RMSE)=0.1 $\mu$VK$^{-1}$ except for UD Bi-2201 (RMSE=0.2 $\mu$VK$^{-1}$). $B_{l}=H_{c2}'$ where
$H_{c2}'$ is the upper critical field $H_{c2}$ at $T_c$. The hole doping $p$ is estimated from the empirical
formula ${T_c}(p) = {T_{c,\max }}[1 - 82.6{(p -0.16)^2}]$ with $T_{c,\max }$=28 K and 90 K in Bi-2201 and
Bi-2212, respectively. $T_c$s are quoted from the literatures identical with Nernst signal. }
\label{ParaNernst}
\begin{center}
\renewcommand\arraystretch{1.5}
\begin{tabularx}{\textwidth}{ >{\setlength{\hsize}{1.3\hsize}}X     >{\setlength{\hsize}{0.7\hsize}}X >{\setlength{\hsize}{0.7\hsize}}X >{\setlength{\hsize}{1\hsize}}X >{\setlength{\hsize}{1\hsize}}X >{\setlength{\hsize}{1\hsize}}X >{\setlength{\hsize}{1\hsize}}X
>{\setlength{\hsize}{1.3\hsize}}X}
\hline
\hline
Sample     & $p$  & $T_c$ (K)     & $T_v$ (K) &$B_0$ (T) &$B_p$ (T) & $B_{l}$ (T) &b   \\
\hline
OP Bi-2201 &\textit{0.16} &\textit{28} &\textit{71.5} &\textit{0} &\textit{2.5} &\textit{48.0} &\textit{0.27}    \\
\hline
\hline
UD Bi-2201  &0.077 &12     &64.9$\pm$5.1    &0.5$\pm$0.5     &0.3$\pm$0.1  &63.5$\pm$5.9     &0.672$\pm$0.131       \\
\hline
OP Bi-2201  &0.16  &28     &65.4$\pm$0.2    &0.1$\pm$0.1     &2.9$\pm$0.1   &50.4$\pm$1.0       &0.639$\pm$0.032       \\
\hline
OD Bi-2201  &0.21  &22     &46.0$\pm$1.5    &0.6$\pm$0.6     &2.0$\pm$0.2     &38.6$\pm$3.3       &0.310$\pm$0.123       \\
\hline
UD Bi-2212  &0.087 &50     &105$\pm$7.5    &0.8$\pm$0.8     &9.1$\pm$2   &182$\pm$17      &0.490$\pm$0.062       \\
\hline
OP Bi-2212  &0.16  &90     &107$\pm$1.6     &19.8$\pm$1.9  &85.8$\pm$8.5   &44.2$\pm$9.9     &0.186$\pm$0.066       \\
\hline
OD Bi-2212  &0.22  &65     &77.3$\pm$2.1    &3.2$\pm$2.1  &37.9$\pm$4.8   &54.7$\pm$16.8     &0.181$\pm$0.18       \\
\hline
\hline
\end{tabularx}
\end{center}
\end{table*}

\begin{figure*}
  \centering
  \includegraphics[width=16cm]{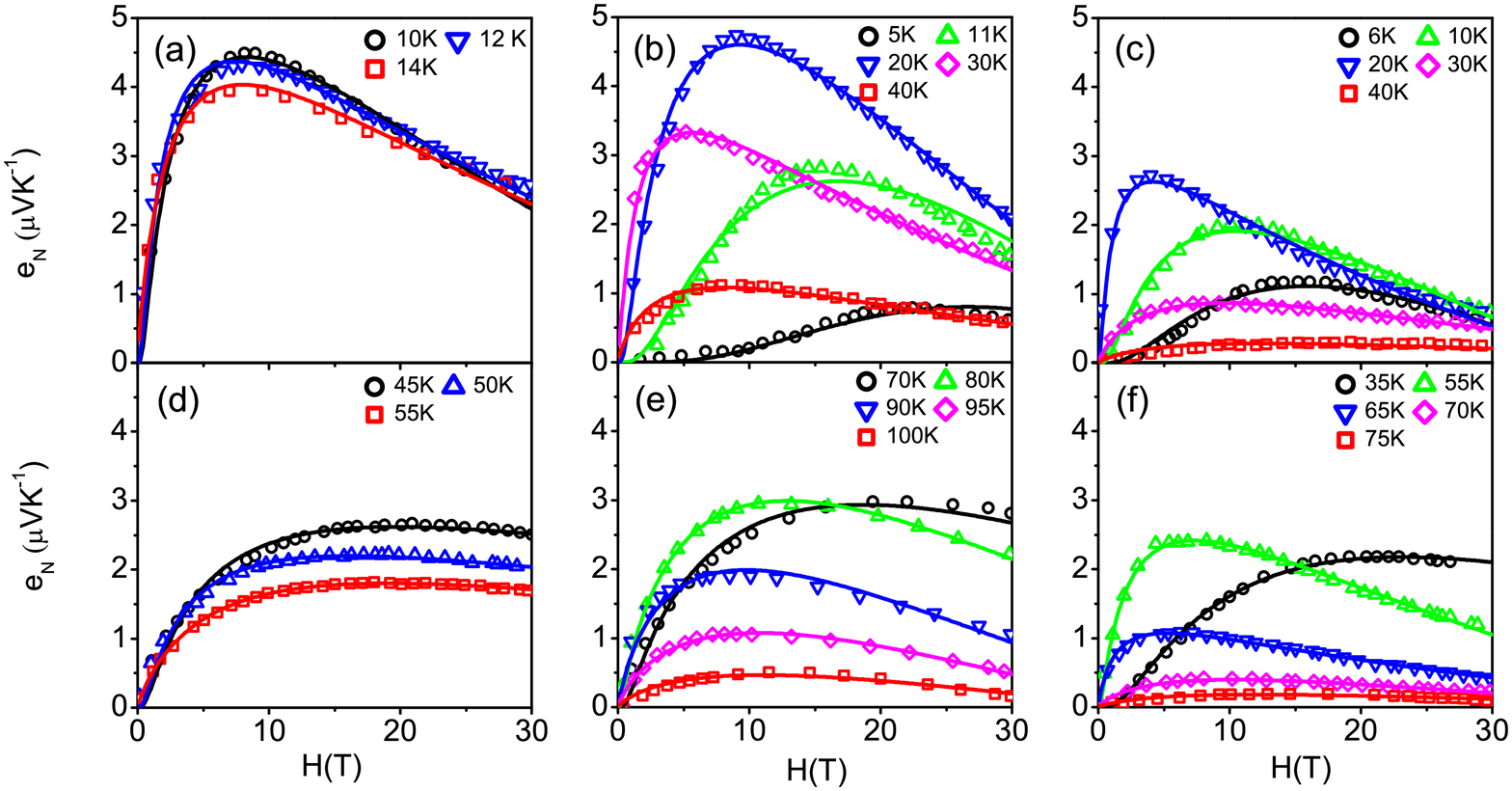}\\
  \caption{Comparison between predictions (solid lines) via Eq. (\ref{eN}) and experimental data (symbols)
of six samples in Bi-2201 \cite{Wang2006} and Bi-2212 \cite{Wang2003,Ong2008}, using parameters
 determined by the FT procedure.  (a), (b), (c) are UD, OP and OD samples in Bi-2201.
  (d), (e), (f) are UD, OP and OD samples in Bi-2212. In UD samples , the characteristic field scale increases at high temperature (e.g., above 20K in UD Bi-2201), which is
often identified as quasiparticle signal \cite{Wang2004}; we thus only deal with the curves at low temperature in these samples.}
  \label{eNcomp}
\end{figure*}

Wang et al. have made systematical measurement of Nernst signal in cuprates superconductor, see Fig.
\ref{eNcomp} \cite{Wang2001,Wang2003,Wang2006}. These observations reveal a  characteristic tilted-hill
profile of vortex-Nernst signal. At low $T$ and $H$, the melting field $H_m$ as well as an exponentially small signal indicate a typical behavior of the TAFF. On the other hand, a linear decay in strong field regime due
to vortex-vortex interaction dominates in $\eta$  is also found. Between these two regimes, a peak appears and moves towards low fields at first and then moves back on the curve $e_N$ vs $H$ when $T$ increases. The peak
vanishes at a characteristic temperature, which called the upper critical temperature $T_v$ in this present
work.

The abundant properties of Nernst signal enable us to determine a set of physical parameters. Comparing to the  magnetoresistence $\rho$ in Eq. (\ref{Resistivity}), the Nernst signal comprises extra complexity
from $S_\phi$ which introduces two $T$-dependent parameters $e_T$ and $H_{c2}$ in Eq. (\ref{eN}). However,
the method to determine the parameters is still the same, a two-step approach (see Appendix \ref{eNpara}).
This two-step approach is applied to various sample of Bi-2201 and Bi-2212 in the present work. In Appendix \ref{eNpara}, OP Bi-2201 is taken as a example to show the specific procedure.

The accurate agreements between the predictions and data are shown in Fig. \ref{eNcomp}, which yields a precise determination of the parameters shown in Table \ref{ParaNernst}. What is important is that the precise agreement extends to the whole range of $H$, $T$ and doping in both monolayer (Bi-2201) and bilayer (Bi-2212) samples. The minimum of root mean square error (RMSE) between predictions and data is
less than 0.1 $\mu$VK$^{-1}$ in most figures, excepts in Fig. \ref{eNcomp}(a) where
RMSE=0.152 $\mu$VK$^{-1}$. This is considered to be remarkable, supporting the validity of this
vortex-fluid model and the reliability of the parameter values.
These parameter values allow us to predict several important physical quantities such as
superfluid  density, upper critical magnetic field and temperature. In the following, we will show that most parameters (especially, $T_v$, $H_{c2}$ and $B_p$) can be reliably determined from the experimental data.

\begin{figure}
  \centering
  \includegraphics[width=7.1cm]{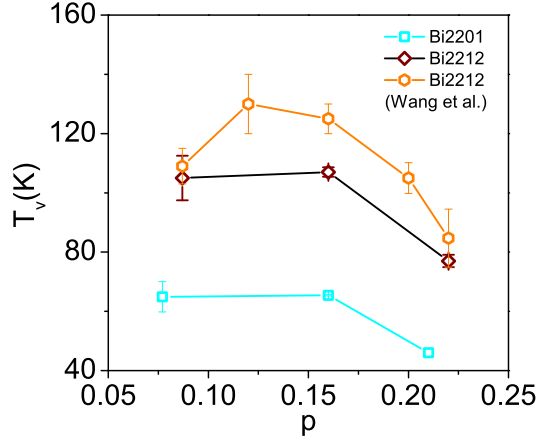}\\
  \caption{Doping dependence of upper critical temperature $T_v$ determined from Nernst signals. The cyan open squares and wine open diamond are data of Bi-2201 and Bi-2212, respectively. The orange open hexagons are data determined by Wang et al. \cite{Wang2006}, and are used for comparison.}
  \label{Tv}
\end{figure}

As discussed in the last section of magnetoresistence, $T_v$ can also be determined for the Nernst signal reliably.  According to the data of most samples, $T_v$ can be determined roughly from the vanishing point of the peak from the curves near zero, and then subjected to a fine tuning.  For example, in OP Bi-2201, the values of $T_v$ from RE and FT are also very close (with a deviation of 8.5$\%$, see, Table \ref{ParaNernst}). In addition, in OP
Bi-2212, we have determined $T_v=107\pm1.6$ K from $e_N$, which is consistent not only with the result (also 107K) from convergent point of $\rho$ (see Sec. \ref{FFR}), but also with the data from Ri et al. \cite{Ri1994} These results strongly support the physical existence of $T_v$ \cite{Chang2012}.
Our final determined $T_v$ for Bi-2201 and Bi-2212 are shown in Fig. \ref{Tv}. Note that the Nernst data at high temperature is laking in UD samples, so its $T_v$ determination has larger uncertainty. The doping
dependence of $T_v$ for Bi-2201 and Bi-2212 are very similar, increasing slowly in the underdoped
regime and decreasing quickly in the overdoped regime. For example, $T_v$ increases from 105$\pm$7.5 K
(UD) to 107$\pm$ 1.6 K (OP), and then decreases to 77.3$\pm$2.1 K (OD) for Bi-2212. Wang et al. defined $T_v$ as the onset temperature where the signal deviates from a background linear $T$ dependence at a given magnetic field \cite{Wang2006}, see orange open hexagons in Fig. \ref{Tv}. Careful examination reveals that in this case, their $T_v$ characterizes a transition to a negative Nernst signal state, while the ours consistently to a zero Nernst signal. Since the negative Nernst signal is beyond the scope of the present work, we consider that our $T_v$ yields a lower bound of upper critical temperature.

The contribution to $\eta$ of vortex-vortex collisions and pinning effect  is characterized by
$f=B\exp[(T_v/T-1)\sqrt{B_p/B}]$(below $T_c$), whose minimum at $T_c$ is $f_{min}=(e^2/4)(T_v/T_c-1)^2 B_p$.
Thus, the values of $T_v/T_c$ and $B_p$ dominate the determination of $B_0$. In OP and OD Bi-2201, UD Bi-2201
and UD Bi-2212, $T_v/T_c>2$, thus  $f_{min}$ is big. Therefore, $B_0\approx0$ is found in these systems. In
contrast, in OP and OD Bi-2212, $B_0$ is comparable to $f_{min}$ (5.6 in OP and 2.4 in OD sample), thus $B_0$
is determined to be a finite value 19.8 T and 3.2 T, respectively.

In the present model, $B_p$ reveals itself as the melting field $H_m$ according to Eq. (\ref{Bp}) in the TAFF
regime. In OP Bi-2201, $B_p$ is found to be stable, varying only 16$\%$ from RE and FT. The resulted
uncertainty of $B_{p}$ is less than or equal to $\pm$10$\%$ for all samples, as shown in Table \ref{ParaNernst}. This reliability of $B_p$ can also be verified from the well described TAFF regime  in OD Bi-2212, OP and OD Bi-2201, as shown in Fig. \ref{eNcomp}. On the other hand, the data at low temperature in UD Bi-2201, UD and OP Bi-2212 are lacking, $B_p$ in these samples requires further validation.

The  determination of $b$ in various samples under different doping is carried out here for the first time. In Bi-2201 and Bi-2212, $b$ decreases monotonically from $0.672$ to $0.310$, and $0.490$ to $0.181$ respectively, with increasing doping. The uncertainties of $b$ in UD Bi-2212, UD and OP Bi-2201 are less than $\pm$11$\%$,
thus the determination of $b$ for these samples is reliable. On the other hand, the uncertainties of $b$ in OD Bi-2201, OP and OD Bi-2212 are bigger than 30$\%$. The reason is understandable: the signals above $T_c$ in
these samples are so small that the relative errors are big. However, the result $b$ = 0.186$\pm$0.066
in OP Bi-2212 is very close to the results determined from magnetoresistance: 0.183 in Bi-2212 ($T_c=84.7$ K)
\cite{Martin1989} and 0.229 in TBCCO ($T_c=99.0$ K) \cite{Kim1989}. From RE to FT in OP Bi-2201, $b$ increases markedly. The reason is that $n_s$ at $T>T_c$ is over estimated with the linear model $n_s=n_{s0}(1-T/T_v)$ in RE, which is balanced in the FT. It seems that the determination of $b$ and $n_s$ are entangled; fortunately,
$b$ can be reliably determined from resistivity, and thus, we propose to perform the resistivity and Nernst
signal measurement simultaneously in one sample, so as to achieve a precise determination of these two
important parameters.

\subsubsection{Superfluid density}

\begin{table*}[tph]
\caption{Characteristic resistivity $\rho_0$, hole density $n_h$ and the characteristic density $n_{s0}$  in
Bi-2201 and Bi-2212. $\rho_0$ is estimated with Eq. (\ref{rhovsp}). Hole density is calculated by
$n_h=n*p*(a_0*b_0)^{-1}$ with the layer number $n$ and lattice constants $a_0$ and $b_0$ in the Cu-O plane,
$a_0=5.36$ \AA\ and $b_0=5.37$ \AA\ in Bi-2201 and $a_0=5.41$ \AA\ and $b_0=5.42$ \AA\ \cite{Zhou1999}. ${n_{s0}}$ is the fitting parameter in Fig. \ref{nscollapse}.}
\label{nhns}
\begin{center}
\renewcommand\arraystretch{1.5}
\begin{tabularx}{\textwidth}{ >{\setlength{\hsize}{1.3\hsize}}X     >{\setlength{\hsize}{0.95\hsize}}X >{\setlength{\hsize}{0.95\hsize}}X >{\setlength{\hsize}{0.95\hsize}}X >{\setlength{\hsize}{0.95\hsize}}X >{\setlength{\hsize}{0.95\hsize}}X >{\setlength{\hsize}{0.95\hsize}}X}
\hline
\hline
Sample        &UD Bi-2201   & OP Bi-2201  & OD Bi-2201  &UD Bi-2212 &OP Bi-2212 &OD Bi-2212 \\
\hline
$\rho_0$ $(${m$\Omega$cm})        &0.41  &0.20  &0.15  &0.19  &0.11  &0.076  \\
\hline
${n_h}$ $({10^{13}}${cm$^{-2}$}) &2.67  &5.56  &7.33  &5.91  &10.9  &14.9  \\
\hline
$n_{s0}$ $({10^{13}}${cm$^{-2}$})            &0.21 &1.40 &0.68 &1.53 &32.0 &9.71\\
\hline
$n_{s0}/{n_h}$     &0.080 &0.25 &0.093 &0.26 &2.93 &0.652  \\
\hline
\hline
\end{tabularx}
\end{center}
\end{table*}

Eq. (\ref{eT}) yields an expression for superfluid density ${n_s}$:
\begin{eqnarray}
{n_s} = {\phi_0 c_0\over C_1 c\rho_0}T e_T.
\label{ns}
\end{eqnarray}
where ${c_0} = 24.6$ \AA for Bi-2201, $30.9$ \AA for Bi-2212. In order to extract ${n_s}$, $\rho_0$ should be
estimated. Since $\rho$ for the Nernst samples is not available in literature, $\rho_0$ needs to be estimated
from   $\rho$ of other (similar) samples with the same $T_c$s. According to Eq. (\ref{Resistivity}), $\rho_0$
is also the resistivity of holes in normal fluid. Based on the Drude model of normal metal, the conductivity
is proportional to hole concentration $p$, thus
\begin{eqnarray}
\rho\propto 1/p.
\label{rhovsp}
\end{eqnarray}
This linear dependence is found to be approximatively valid in Bi-2212 \cite{Usui2014}, hence $\rho_0$ can be
determined for each sample once a value of $\rho_0$ at some doping is obtained. The results are shown in Table \ref{nhns}, using the data $\rho_0\approx0.11$ m$\Omega$cm in OP (p=0.16) $\rm Bi_{2.1}Sr_{1.9}CaCu_2O_{8+\delta}$  \cite{Usui2014} and $\rho_0\approx0.13$ m$\Omega$cm in the OD (p=0.24)
Bi-2201 \cite{Ando1996}. Strictly speaking, the concentration of oxygen deficiency in the CuO$_2$ plane are
different in each sample so that the absolute deviation of this rough estimation of $\rho_0$ and then $n_s$
may become large, sometimes over $100\%$ \cite{Ando1996}. However, the magnitude, relative value in various
temperature of $n_s$ determined by the present strategy are reliable.

\begin{figure}
  \centering
  \includegraphics[width=7cm]{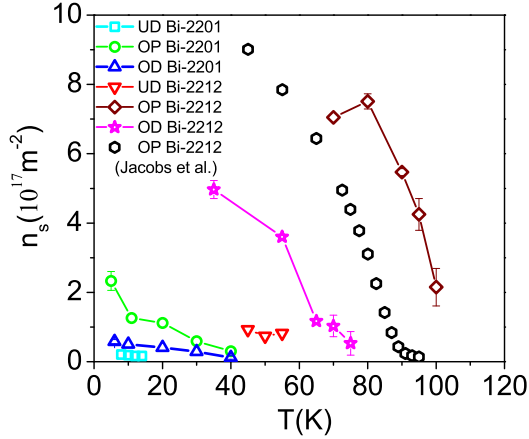}\\
  \caption{Temperature dependence of superfluid density ${n_s}$ determined from $e_T$ with Eq. (\ref{ns}) in
  six samples of Bi-2201 \cite{Wang2006} and Bi-2212 \cite{Wang2003,Ong2008}. Note that we add a set of $n_s$, in solid and black hexagons, from penetration depth $\lambda$ \cite{Jacobs1995} in OP Bi-2212 ($T_c=91$ K,
  $n_{s}(0)=1.29\times10^{14}$ cm$^{-2}$). The errors of $n_s$ are small for most data.}
  \label{nsT}
\end{figure}

${n_s}$ of six samples in Bi-2201 and Bi-2212 obtained from Eq. (\ref{ns}) are shown in Fig. \ref{nsT}. There
are two distinct features: a nearly linear temperature dependence in most $T$ regime below $T_v$ and a finite
$n_s$ above $T_c$. The linear $T$ dependence of $n_s$ at low temperature has been widely observed in HTSC
\cite{Panagopoulos1998}. In OP Bi-2212, this linearity is found as well as in microwave penetration depth
\cite{Jacobs1995} (see, open hexagon in Fig. \ref{nsT}) and  in magnetization $M$ \cite{Waldmann1996}. In the
standard BCS theory of d-wave superconductor,  an order parameter with nodes in the gap leads naturally to a
linear $T$ dependence at low temperature. Thus, the linearity is considered to be an evidence supporting the
d-wave symmetry for cuprates superconductor. It is noteworthy that in the BCS theory, the decrease of $n_s$
with $T$ describes the depairing effect due to thermal fluctuations. It is intriguing to question whether the
observed linear $T$ dependence here is indeed resulted from dephasing due to thermal fluctuations and how it
can be derived quantitatively.

It is important to note that the $T$ dependence of $n_s$ determined from $e_N$ seems to be universal.  As shown in Fig. \ref{nscollapse}, $n_s$ collapses on a straight line $n_s=n_{s0}(1-T/T_v)$, where $n_{s0}$ is
derived from the fitting parameter shown in Table \ref{nhns}. A good indication is that $n_{s0}$ is close to
the hole density ${n_h}$ in most samples (except UD samples), see $n_{s0}$ in Table \ref{nhns}, which proves
that our model is sound. For UD samples, the lacking of $e_N$ data at low temperature results in an
underestimate of $B_p$ (see Table \ref{ParaNernst}, $B_p$ for UD samples are only 1/10 of the $B_p$ for OP
samples with same layers), which is the cause for an underestimate of $n_s$ and $n_{s0}$. In OP Bi-2212,
$n_{s0}$ is 2.9 times of $n_h$, which may be caused by an underestimate of $\rho_0$ in Eq. ({\ref{ns}}).
Meanwhile, an anomalous decrease at low temperature (see Fig. \ref{nsT}) is believed to be caused by an
inhomogeneous distribution of disorder at low temperature, which contributes to extra pinning effect, not
considered in our model. This inhomogeneity is wiped out by thermal fluctuations at higher temperature.
The non-constant values of  $n_{s0}/n_h$ reflect doping and layer effect. Also an effective mass effect is
neglected by setting the hole effective mass to be bare electron mass in Eq. (\ref{Sphi2}).

Despite all these effects, the observed universality is impressive. It indicates that the linear $T$
dependence of $n_s$ is robust. In addition, the $n_s$ presently derived from $e_N$ signal is reliable.
Furthermore, the vanishing of $n_s$ at $T_v$ but not $T_c$ is very indicative of the nature of the pseudogap.
Anderson proposed that the $n_s$ derived from $e_N$ above $T_c$ is the local superfluid density which flows
within interstitial puddles between vortex cores \cite{Anderson2007,Wang2006}, which is supposedly different
from global superfluid density. The latter, when measured in penetration depth \cite{Jacobs1995} (see, open
hexagon in Fig. \ref{nscollapse}), must be suppressed by vortex tangles at large scale thus decays quickly
above $T_c$. However, $n_s$ determined from $e_N$ in the present work is finite and larger than the previous
measurements (i.e., decays slowly) above $T_c$, indicating that we have obtained a true local superfluid
density.

\begin{figure}
  \centering
  \includegraphics[width=7cm]{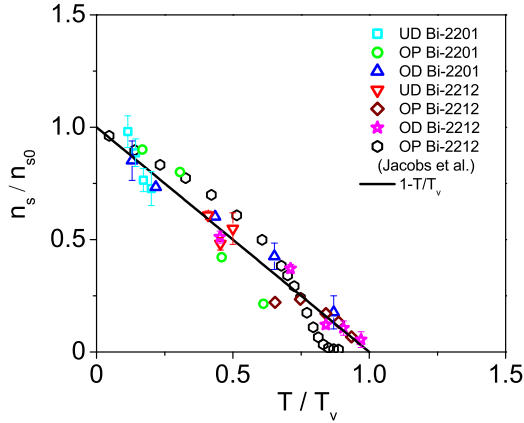}\\
  \caption{The normalized superfluid density ${n_s}/{n_{s0}}$ vs $T/{T_v}$ for
   six samples in Bi-2201 and Bi-2212, where $n_{s0}$ is shown in Table \ref{nhns}. The data are the same as the one used in Fig. \ref{nsT}. A linear law $1 - T/T_v$ is obtained. The open and black hexagons are the data $n_s/n_{s}(0)$ derived from microwave penetration depth ($n_{s}(0)=1.29\times10^{14}$ cm$^{-2}$) \cite{Jacobs1995}. }
  \label{nscollapse}
\end{figure}

\subsubsection{Upper critical field}

\begin{figure}
  \centering
  \includegraphics[width=7cm]{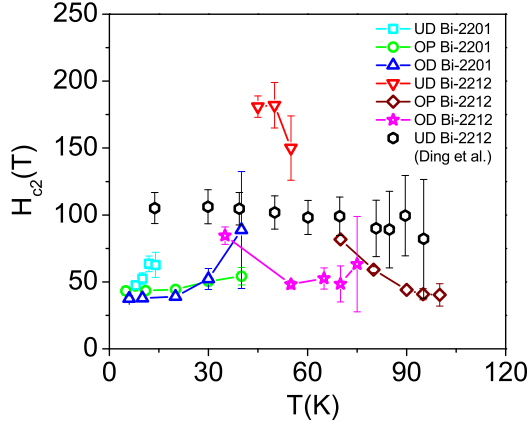}\\
  \caption{Temperature dependence of  ${H_{c2}}$ determined from $e_N$ via Eq. (\ref{eN}) in six samples of
  Bi-2201 \cite{Wang2006} and Bi-2212 \cite{Wang2003,Ong2008}. Since $H_{c2}$ is the typical characteristic
  of $e_N$ vs $B$, the errors for most points are small. While for OD Bi-2201 and Bi-2212, the small slopes near $T_v$ result in big uncertainties of $H_{c2}$. $H_{c2}$ in UD Bi-2212 ($T_c=80$ K), determined from gap amplitude \cite{Ding2001} with Eq. (\ref{Hc2gap}), are shown by open hexagons for comparison.}
  \label{Hc2T}
\end{figure}

An interesting quantity which attracts a wide attention is the upper critical field  $H_{c2}$
\cite{Wang2003,Chang2012}, beyond which the Nernst effect disappears. $H_{c2}$ is too high to measure in
cuprates, and reported estimates show sharp contradictions \cite{Chang2012}. Since $H_{c2}$ represents a
crossover from a vortex fluid to the normal state, Wang et al. \cite{Wang2003,Wang2006} defined $H_{c2}$ as
the field where $e_N$ versus $B$ extrapolates to zero; this yields two conclusions that $H_{c2}$ in HTSC is
almost $T$ independent and $H_{c2}$ decreases as hole concentration increases, which are consistent with $T$
and $p$ dependence of the gap amplitude $\Delta_0$ measured by ARPES \cite{Ding2001}. On the other hand, Chang et al. treated $B^m=H_{c2}\ln(T/T_c)$ as a new definition of $H_{c2}$ based on Gaussian-fluctuations theory
\cite{Chang2012}, where $B^m$ is the peak field. When doping increase, a non-monotonic variation of $H_{c2}$
is observed in Eu-LSCO. Hopefully, these two empirical determinations are to be clarified based on an
accurate quantification of $e_N$ data. Our model enables one to make use of the data in a wider range of $T$
and $H$ to determine ${H_{c2}}$, thus may be considered as a more sophisticated version of the simple linear
extrapolation of Wang et al.,  leading to a more precise determination of ${H_{c2}}$, especially when the
measured profile is imperfect.

Using the method introduced in Sec. \ref{eNpara}, $T$ dependence of ${H_{c2}}$ is determined and shown in Fig. \ref{Hc2T}. In OP Bi-2201, $H_{c2}$ is in $[44,53.6]$ T by the RE, which is used as the range for performing FT, as shown in Table \ref{TDpara}. The final results (see Fig. \ref{Hc2T}) of FT is close to the rough
estimates. In Fig. \ref{Hc2T}, most curves show a slow variation near $T_c$, which are generally consistent
with the results of Wang et al., in which $H_{c2}$ fluctuates with $\pm4$ T around the mean value 48 T in OP
Bi-2201 \cite{Wang2006}.  The nearly $T$ independence of $H_{c2}$ is closely related to the gap amplitude
$\Delta_0$  observed in ARPES \cite{Ding2001}. An uncertainty-principle arguments results in a relation
between $\Delta_0$ and  Pippard coherence length $\xi_P=\hbar v_F/a\Delta_0$, where the Fermi velocity  $v_F$
and numerical constant $a=1.5$   can be estimated as $1.78\times10^5$ ms$^{-1}$ and $a=1.5$, respectively
\cite{Wang2003}.  Suppose $\xi\approx\xi_P$, $H_{c2}$ can be calculated as
\begin{eqnarray}
H_{c2}=C_2\Delta_0^2,
\label{Hc2gap}
\end{eqnarray}
where $C_2={\phi_0a^2/{2\pi\hbar^2v_F^2}}$. The results for UD Bi-2212 \cite{Ding2001} ($T_c=80$ K) are shown
as open hexagons in Fig. \ref{Hc2T}, which reveals a distinct $T$ independence.

Note that the observed linear increase of $H_{c2}$ above $T_c$ in UD and OD Bi-2201 is a result of small
negative slope of $e_N$ in high fields, see Fig. \ref{eNcomp}. In OD Bi-2201, however, one finds a nonlinear
tail in high fields, which may be generated by amplitude fluctuations, and which appears commonly in Nernst
and Ettingshausen signal in conventional superconductors, YBCO \cite{Ri1994} and electron-doped
(Nd$_{2-x}$Ce$_x$CuO$_{4-y}$) HTSC \cite{Wang2006}. In UD Bi-2201, $e_N$ reaches a platform in high fields and the decline regime is invisible at $T$ much higher than $T_c$, thus the field scale of $e_N$ increase
anomalously \cite{Wang2004}. This anomalous increase prevents ${H_{c2}}$ from being accurately estimated at
high $T$ in UD sample, and is interpreted as a contribution of incoherent phase with a strong pairing
potential (thus large field scale) \cite{Wang2004}. In our opinion, once the contribution of quasiparticle and amplitude fluctuations are properly described, more precise determination $H_{c2}$ above $T_c$ in Bi-2201 can
be achieved. It is worthwhile to mention that, in Bi-2201 samples, the typical Nernst signal (tilted-profile)
only deforms at $T$ far above $T_c$, which means that the vortex-fluid model is valid in most of the physical
domain. Therefore, the main results of Bi-2201 in the present work are reliable.

Since the quasiparticle contribution in Bi-2212 is very small, the determination of $H_{c2}$ is more precise
than in Bi-2201. In Fig. \ref{Hc2T}, ${H_{c2}}$ varies slowly near $T_c$ in OP and OD Bi-2212. This is
consistent with nearly $T$ independent gap amplitude ${\Delta _0}$ observed in ARPES \cite{Ding2001}.  A
interesting behavior is the approximately linear decrease near ${T_c}$  in OP Bi-2212. This is very similar as
the case in electron doped cuprates Nd$_{2-x}$Ce$_x$CuO$_4$ \cite{Wang2003} and conventional superconductor at $T_c$. Whether it is an indication of decrease of the coherence length or a effect of quasiparticle signal
need further investigation when data at higher temperature are available.

It is worthwhile to discuss a special  $H_{c2}$ at $T_c$, denoted by ${H_{c2}'}$, which reveals a major
feature of HTSC, different from conventional linear dependence ${H_{c2}}\sim({T_c}-T)$, vanishing at $T_c$. In the RE (linear extrapolation in high fields) in Bi-2212, ${H_{c2}'}$ decreases from 137$\pm$23 T, 52$\pm$2.1 T to 43$\pm$3.3 T as the doping increases, close to the results of Wang et al. \cite{Wang2003} who reported
${H_{c2}'}$ being 144, 67 and 50 T, respectively, for three dopings. In the FT, as shown in Table \ref{ParaNernst}, ${H_{c2}'}$ are 182$\pm$17, 44.2$\pm$9.9 to 54.7$\pm$16.8 T in Bi-2212. The significant
difference between RE and FT reveals that the determination of ${H_{c2}'}$ is error-sensitive in Bi-2212 since  the field scale in Bi-2212 is too large to be reachable by the $e_N$ data. Thus, the conclusion of doping
dependence of ${H_{c2}'}$ in Bi-2212 requires further verification. Nevertheless, the overall $T$ dependence
of $H_{c2}$ shown in Fig. \ref{Hc2T} is reliable. The figure shows that, at same $T$, $H_{c2}$ in OP sample is always higher that $H_{c2}$ in OD sample. Finally, in both cases of Bi-2201 and Bi-2212, we reach the
conclusion that ${H_{c2}}$ decrease with increasing doping.

\section{Discussion and Conclusion}\label{DisCon}

After the Nernst effect in pseudogap state and the supersolidity in solid helium-4 are discovered, Anderson
insightfully proposed that they are vortex liquids and the dynamics are controlled by thermally excited vortex
tangles \cite{Anderson2007}. The transport physics of such vortex fluid is interesting, but complicated due to
multiple mechanisms in play, and hence has not yet been addressed quantitatively. The present phenomenological
model takes upon this challenge, with a quantitative formalism involving all relevant processes such as
thermal and magnetic vortex activation, impurity scattering, vortex-vortex collisions and flux pinning, etc.
The major conclusion of the work is that Nernst signal over a wide range of $T$ and $B$ can be quantitatively
described by vortex-fluid model, taking sound parameter values. The theoretical predictions and experimental
data of Nernst effect agree remarkably well over the entire range of $T$ and $B$ experimentally reported,
which is rare in HTSC studies. A significant outcome is the demonstration of the ability to extract the values of eight parameters, with a two-step procedure, from the experimental data of magnetoresistance and Nernst
effect, despite the complicated analytic form.

The current model is readily applicable to analyze any new layered or two dimensional superconductor. Once the magnetoresistance $\rho$ and Nernst signal $e_N$ data of a new sample are available, the rough estimate can be
conducted first using asymptotic analysis at large and small limits of $T$ and $B$. Note that for
electron-doped cuprates, some UD hole-doped cuprates \cite{Chang2012}, and even iron-based superconductor
\cite{Pourret2011}, the contributions of quasiparticles should be subtracted with a linear or nearly linear
model \cite{Wang2006,Wang2004}. The procedure of rough estimate can be summarized as following. Step one,
intuitively estimate the upper critical temperature $T_v$ and characteristic resistivity $\rho_0$ from the
convergent (or saturated) regime of $\rho$. Step two, carry out a magnitude analysis of the maximum $e_N$
signal with Eq. (\ref{ec}), and predict the characteristic superfluid density $n_{s0}$ if the carrier
effective mass is available (e.g., for Fe$_{1+y}$Te$_{0.6}$Se$_{0.4}$, $m_c=29$ $m_e$ \cite{Pourret2011}). The
comparison between $n_{s0}$ and carrier density in the sample allows to judge the validity of the vortex-fluid scenario. Step three, determine $H_{c2}$ (thus $B_l$) from the linear extrapolation of $e_N$ in strong fields
according to Eq. (\ref{eNtoHc2}). Sometimes, linear regime of $e_N$ in strong fields is not measured (e.g.,
for Fe$_{1+y}$Te$_{0.6}$Se$_{0.4}$ \cite{Pourret2011}), then use the values from other measurements of similar sample \cite{Fang2010} so that the determination of $n_s$ in the final step can still be achieved. Step four,
use the data at low $T$ of $\rho$ (or $e_N$) (with significant pinning effect) to determine $B_p$ with  Eq.
(\ref{Bpfitrho}) (or Eq. (\ref{BpfiteN})). Step five, determine $B_0$ from the linear field dependence of
$\rho$ or $e_N$ in low fields at $T_c$. Step six, determine the Kosterlitz coefficient $b$ from the sharp
increase near $T_c$ of $\rho$ in zero field with Eq. (\ref{rhozero}). Finally, estimate the $T$ dependent
local superfluid density $n_s$ by the calculation of the peak of $e_N$ vs $B$ with Eq. (\ref{eNm}) and
Eq. (\ref{ns}). By the way, although it is better to obtain measurements of magnetoresistance and Nernst
signal simultaneously and in wide $T$ and $B$ range, there are other ways to complement the data from the
literatures using our vortex-fluid model.

Details of the above procedure are presented in the Appendix \ref{ParaDeter}, and the comparison of presently
obtained values to prior measurements shows that our determination is reliable and can be regarded as an
improved version of former intuitive determinations, such as that by Wang at al. \cite{Wang2006}
These precisely determined parameter values characterize the vortex motions, and enable us to interpret the
characteristic tilted-hill profile and its peak shift in terms of the variation of vortex density and transport velocity, which will be reported elsewhere.

It is interesting that the newly derived parameter values allow new predictions of the local superfluid
density within interstitial puddles between vortex cores, which reveals a universally linear $T$ dependence
below $T_v$. These results raise new questions related to transport properties in HTSC. For example, how to
explain the origin of the linear $T$ dependence of local superfluid density, especially above $T_c$? The
difference between local (i.e., microscopic) superfluid density from Nernst signal and the global (i.e.,
macroscopic) superfluid density from conventional penetration depth needs further clarification; the latter is an average over phase fluctuations \cite{Anderson2007}. Another question concerns the definition of the
critical state between vortex fluid and normal state.  Above $T_c$, thermal and magnetic vortices may
interact, which will complicate further the description. One important question is whether the classical
relation $H_{c2}=\phi_0/2\pi\xi^2$ still hold, and whether the increase of vortex density (due to thermal
vortices) results in a redistribution of vortex cores (thus a modification of geometric factor 2$\pi$)?  Since $T_v$, $H_{c2}$ and $n_T$ (thermal vortices density) are all quantitatively determined in present work, these
questions can be addressed within the present framework.

The most important construction in our vortex-fluid model is Eq. (\ref{eta}) of damping viscosity $\eta$, which is an extension of the Bardeen-Stephen model to include the contributions of flux pinning and vortex-vortex
collisions, all of which are necessary in HTSC. Note that our estimate of $\eta$ from magnetoresistance data
is two order of magnitude higher than Anderson's original proposal based on thermal and quantum fluctuations.
Interestingly, a recent analysis of diamagnetism and conductivity with vortex-plasma model reported a rather
small viscosity (one to two orders of magnitude smaller) in a sample of La$_{1.905}$Sr$_{0.095}$CuO$_{4}$
($T_c=23.5$ K) \cite{Bilbro2011}, but a careful analysis reveals an inconsistency with magnetoresistance
data in a similar sample of La$_{1.92}$Sr$_{0.08}$CuO$_{4}$ ($T_c=28$ K). If we assume that both samples have
similar viscosity ($\sim1\times10^{-9}$ gs$^{-1}$cm$^{-1}$) at $T=23.5$ K as  \cite{Bilbro2011} predicted,
then substitute it into (widely accepted) Eq. (\ref{Resistivity-formula}) (at $H=2$ T), $\rho$ is found to be
40 times higher than the data. \cite{Suzuki1991}, indicating that the viscosity predicted in
\cite{Bilbro2011} is too small. In fact, some effects of the short-range coherent surface current and
quasiparticles may compensate the effect of thermal vortices in diamagnetic response near $T_c$, leading to
substantial underestimate of its density, hence the magnitude of the viscosity. We consider that as an open
question to be investigated in the future.

More interestingly, the discrepancy of Anderson's model suggests the existence of anomalous transport of
vortices in HTSC beyond mere quantum and thermal fluctuations. In classic fluid, vortex motions often display
chaotic behavior, so exhibiting enhanced dissipation and momentum transport. What is the nature of the
transport in the vortex fluid in HTSC, and whether a turbulent-like behavior emerges in the vortex
entanglement (in the presence of pinning)? Is the strengthened damping viscosity generated by complicated
interactions between vortices and with pinning centers, which creates a turbulence involving multiple scales?
There may be exotic answer to these two interesting questions.

In previous studies, vortex motions are believed to originate only from phase fluctuations when amplitude is
frozen, implying that amplitude fluctuations play no role in vortex fluids \cite{Podolsky2007}. Here, we
argue, based on an in-depth consideration of the lifetime $\tau_{GL}$ of fluctuation Cooper pairs (or
amplitude fluctuations) \cite{Larkin2005}, that magnetic field can also drive fluctuation pairs to form a
circular current, thus fluctuation vortex if coherence length is small. A dimensional estimate yields that
${\tau _{GL}} \sim \hbar/{k_B}(T - {T_c}) \sim$10 ps at $T-T_c=$10 K, which is much larger than the cycle
period of the pair near the core, which is ${\tau _c} \sim 2\pi {m_e}{\xi ^2}/\hbar  \sim$ 1 ps when
$\xi\sim$3 nm. The fact that ${\tau _c}\ll{\tau_{GL}}$ implies that fluctuation Cooper pairs  form realistic
circular current in the form of fluctuating magnetic vortex in HTSC. The integrated picture of phase and
amplitude fluctuations is like this, in UD samples where phase fluctuations dominate, the fluid is composed of conventional vortices; in OD samples, the fluctuating vortices play the leading role.

In summary, the present vortex-fluid model accurately describes, for the first time, both the
magnetoresistance and Nernst signal for HTSC over a wide range of $T$ and $B$. This enables us to reliably
extract a series of physical parameters characterizing the vortex motions. These parameters are important to
the study of the physical properties of the HTSC system, possibly for revealing different transport properties of Fe-based from cuprates HTSC. Furthermore, the study of the temperature, doping and layer dependence of
parameters will reveal new physics related to phase fluctuations and the mechanism of strong entanglement of
vortices in pseudogap phase.

\begin{acknowledgments}
We thank L. Yin for many constructive discussion during the early stage of this work. We thank Y. Wang for a
very helpful discussion. This work is partially supported by National Nature Science (China) Fund 11452002.
\end{acknowledgments}

\begin{appendix}

\section{Parameters determination}\label{ParaDeter}

A key result of the present work is to demonstrate that it is possible to derive reliably the values of the model parameters, as we show below.

The determination of model parameters is conducted in two steps. The fist step is a rough estimation (RE),
using asymptotic analysis at high and low limits of $T$ and $B$, where naturally one or two leading mechanisms are dominant. In this case, only one or two parameters appear, so their approximate values can be obtained.
This determination is approximate because empirical data do not go very far in the asymptotic limits; more
often the effects of variation of the multiple parameters are entangled together, so that a global
optimization is necessary to find more accurate description of data. This is then accomplished by the second
step of fine tuning, which define the best set of parameter values by minimizing the error between the
predictions and empirical data.  In other words, the parameters are tuned around the roughly estimated values
to achieve a better fitting of a set of data (multi-curves).

The fine tuning (FT) is realized by a least square fitting process. For example, in the determination with
Nernst signal, we divide the entire set of parameters into groups so that the FT can be conducted
one group after another, with the former results substituted into the later calculation. This iterative
process runs until a local minimum of the error is reached. If there are more than one local minimums, a
comparison between them decides the global minimum, and then the FT ends with a set of precisely
determined parameters. Sometimes, the physically optimal point is not the global minimum but the local
minimum. Therefore, to determine the local minimum, we need to define the range of parameters to be explored,
to be specified in the main text.

The reliability of the parameters depends on the precision of the comparison between predictions and the data, and the uncertainties of the parameters. The later can be expressed by error bar. Once the optimal set of
parameters are obtained, the error bar is determined once the upper boundary of the root mean square error
(RMSE) between the predictions and empirical data reaches a threshold equals to the measurement uncertainty.

\subsection{Determine parameters with flux-flow resistivity}\label{rhopara}

In the resistivity model Eq. (\ref{Resistivity}), there are seven parameters, namely $T_c$, $\rho_0$, $T_v$,
$B_0$, $B_p$, $B_{l}$ and $b$, to be determined by experimental data. $T_c$ can be easily estimated from the
transition temperature of zero resistivity. In this section, we explain how other parameters are determined,
using magnetoresistance data of optimally doped ($p=0.16$) $\rm Bi_{2.1}Sr_{1.9}CaCu_2O_{8+\delta}$
\cite{Usui2014}.

\subsubsection{Rough estimate}

First, let's discuss the RE. In Sec. \ref{FFR}, we assume that the high $T$ limit of thermal vortex density is as dense as magnetic vortices at $H_{c2}'$, thus $B_{l}\approx H_{c2}'$. $H_{c2}'$ can be estimated as $\phi_0/(2\pi\xi^2)$. Thus, $B_{l}\approx\phi_0/(2\pi\xi^2)=68.1$ T as $\xi=22$ \AA\ is measured by STM \cite{Pan2000}.

Near $T_v$, vortex-vortex collisions are dominant in $\eta$ since the density of thermal vortices is dense,
thus $B_0$ can be neglected. Then, $\rho$ can be expressed as
\begin{eqnarray}
\rho\approx\rho_{0}\exp[-({T_v\over T}-1)\sqrt{B_p\over {B+B_T}}],
\label{rhoTv}
\end{eqnarray}
which predicts that $\rho$ increases with $B$ for $T<T_v$ and converge to $\rho_0$ for $T$ approaching $T_v$.
We use the convergent regime of the experimental curves (see Fig. \ref{rhocomp}) to carry out a rough estimate of $T_v$ and $\rho_0$. The convergent regime corresponds to the range where the curves at the highest (17.5 T)
and lowest (0) become close (or overlapped). In practice, we consider the overlap starts where the difference
of the two curves equals the scatters of each curve, which is about 5 $\mu\Omega$cm. Applying this criteria,
we obtain $T_v\in[104,120]$ K and $\rho_0\in[0.098,0.133]$ m$\Omega$cm. The middle values, e.g., $T_v=112$ K
and $\rho_0=0.116$ m$\Omega$cm, are taken to be the roughly estimated values.

$B_p$ is determined from the low $T$ limit (i.e., the TAFF regime) of the resistivity profiles. At $T\ll T_c$, pinning effect dominates and $B_0$ is neglected.  We define $y=\ln(\rho_{0}/\rho)$, $x=T_v/T-1$, then a linear formula is obtained,
\begin{eqnarray}
y=a_0x+b_0,
\label{Bpfitrho}
\end{eqnarray}
where $a_0=\sqrt{B_p/B}$, and $b_0$ contains the contributions from $B_0$ and data error. A least square
fitting of the line yields $a_0\approx 1.85$ so that $B_p\approx 30.8$ T, as shown in Fig. \ref{rho0rho}.

\begin{figure}
  \centering
  \includegraphics[width=7cm]{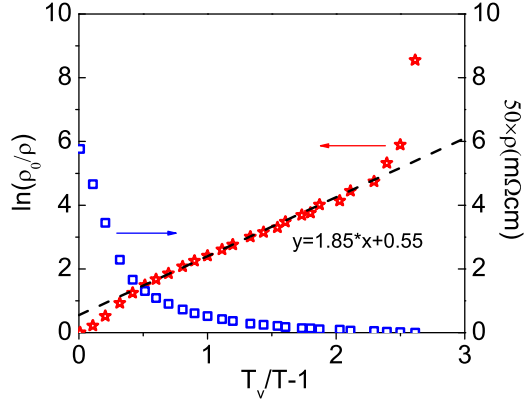}\\
  \caption{Linear fitting of $\ln{(\rho_{0}/\rho)}$ vs $T_v/T-1$ at $B=9$ T. The red open stars are the data of
  $\ln(\rho_{0}/\rho)$. The black dash line is the linear fitting curve. The blue open squares are the data of $\rho$.}
  \label{rho0rho}
\end{figure}

$B_0$ is supposed to dominate in low fields where vortices flow individually in materials with weak pinning.
However, for $\rm Bi_{2.1}Sr_{1.9}CaCu_2O_{8+\delta}$ \cite{Usui2014}, the melting field $H_m\geq17.5$ T
indicates a significant pinning effect, thus $B_0$ is nearly invisible at low temperature. The contribution of vortex-vortex collisions and pinning effect is presented by $f$,
 \begin{eqnarray}
f=B\exp[(T_v/T-1)\sqrt{B_p/B}].
\label{f}
\end{eqnarray}
To suppress this contribution  and avoid the effect of thermal vortices, we propose to determine $B_0$ at
$T_c$ and near $H$', where $H$' is the minimum of $f$ at $T_c$. At $T_c$, the Eq. (\ref{Resistivity}) is
simplified to
\begin{eqnarray}
\rho=\rho_{0}{B\over{B_0+f}}.
\label{rhoHTc}
\end{eqnarray}
Since $T_v$ and $B_p$ are determined, the minimum of $f$ is found to be $f_{min}=(e^2/4)(T_v/T_c-1)^2 B_p$=4.7 T at $H'=0.64$ T.  According to calculation, $f$ is approximately a constant near $H'$, thus  the data at
$H=0.5$T ($\rho=10.9$ $\mu\Omega$cm) and 1 T ($\rho=16.6$ $\mu\Omega$cm) are used in the linear fitting.
Finally, $B_0\approx4$ T is obtained.

$b$ is determined near and above $T_c$ with Kosterlitz model \cite{Kosterlitz1974}. At $T>T_c$, we have the
following relation:
\begin{eqnarray}
B_{0}= (B_i+B_{T})\{\frac{\rho_{0}}{\rho_i}- \exp[t\sqrt{\frac{B_p}{B_i+B_{T}}}]\},
\label{B0}
\end{eqnarray}
where $i=$a,b. Equaling the case a and b, we obtain $B_{T}$. $B_{T}$ calculated from $B_a=0.5$ T and $B_b=9$ T in the range $T \in[90,100]$ K is shown in Fig. \ref{bfitrho}. According to Eq. (\ref{BT}), the Kosterlitz
coefficient $b=0.42$ is fitted from the data at $T\in[90,94]$ K.

\begin{figure}
\centering
\includegraphics[width=7cm]{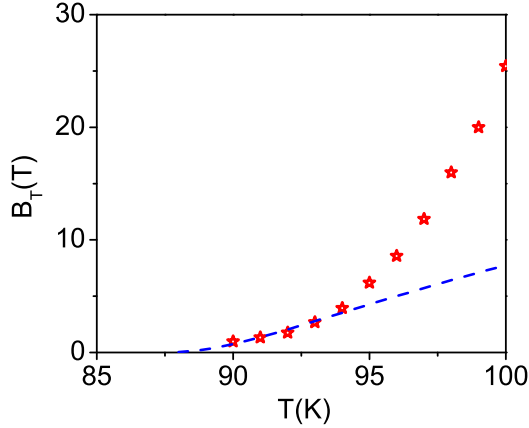}\\
\caption{Determination of $b$ from the fit (blue dashed line) of $B_T$ (red symbols) based on  Kosterlitz model Eq. (\ref{BT}).}\label{bfitrho}
\end{figure}

\subsubsection{Fine tuning}

In this step, we vary the five parameters ($\rho_0$, $T_v$, $B_0$, $B_p$ and $b$) around their roughly
estimated values to get a precise fitting of the experimental curves, while keeping  $B_{l}=68.1$ T unchanged,
following the argument in the beginning of this section. This fine tuning process needs to set up a ranges for each parameter. First, the result of RE already yields $T_v\in[104, 120]$ K and $\rho_0\in[0.098, 0.133]$
m$\Omega$cm. Secondly, $B_0$, $B_p$ and $b$ are set to be positive, as it is evident from their physical
meaning. Then, we treat the empirical data of 5 curves at $H=$0, 0.5, 3, 9 and 17.5 T for $T\leq104$ K  as an
ensemble, and evaluate the RMSE of the predictions from the data, when the five parameters vary in the range
specified. Finally, the smallest RMSE $\approx3.5$ $\mu\Omega$cm is reached and the fitting is shown in Fig.
\ref{rhocomp}.

According to the procedure introduced in the beginning of this section, the error bar can be determined from
the uncertainty of the original data \cite{Usui2014}, which approaches $\pm2.5$ $\mu\Omega$cm. We thus use 5
$\mu\Omega$cm as the upper boundary of the RMSE function. The error bar for any one parameter is calculated
when the others are set to their FT-determined values, and the results are shown in Table \ref{ParaRes}.

\subsection{Determine parameters with Nernst effect} \label{eNpara}

The systematically measured Nernst signal  allows us to do reliable parameter determinations. In this part,
Eq. (\ref{eN}) is applied to the six samples in Bi-2201 and Bi-2212 \cite{Wang2003,Wang2006,Ong2008}. The
process of parameters determination with $e_N$ is similar to the one with $\rho$ but there are some more steps due to the two more $T$ dependent parameters ${e_T}$ and $H_{c2}$. In the following, OP Bi-2201 is taken as an example to show the specific procedures.

\subsubsection{Rough estimate}

In RE, the asymptotic analysis at high and low limits of $T$ and $B$ is also the key strategy. $T_v$ can be
determined from the zero point of the linear extrapolation of the peak at very high temperature. Using
similar linear extrapolation in high fields, $H_{c2}$ is obtained. ${e_T}$ and $B_p$ can be determined from
field dependence of the signal in the TAFF regime. The nearly linear field dependence of $e_N$ near the
minimum of $f$ below and near $T_c$ can be used to determine $B_0$. Meanwhile, $b$ can be determined from the
peaks of the profiles above $T_c$.

On the profile of ${e_N}$ vs $B$, when $T \to {T_v}$,
\begin{eqnarray}
{e_N} \to {e_T}\frac{B}{{{B_0} + B + {B_{T}}}}\ln (\frac{{{H_{c2}}}}{{B}}).
\label{eNTv}
\end{eqnarray}
Since $n_s$ approaches 0 at $T_v$, the peak $e_N^m$ of the profile approaches zero.  Since $e_N^m$ vs $T$ is a nonlinear curve at large $T$ range, only the data (50 K and 65 K) nearest to zero are used in the linear
extrapolation, shown in Fig. \ref{TvHc2}(a). Finally, ${T_v} =71.5$ K is obtained. The uncertainty of the data
is estimated to be $\pm$100 nVK$^{-1}$ (discussed in the end of this section), and the error bar of $T_v$ is
found to be $\pm6.2$, thus ${T_v} = 71.5\pm6.2$ K.

\begin{figure}
  \centering
  \includegraphics[width=6cm]{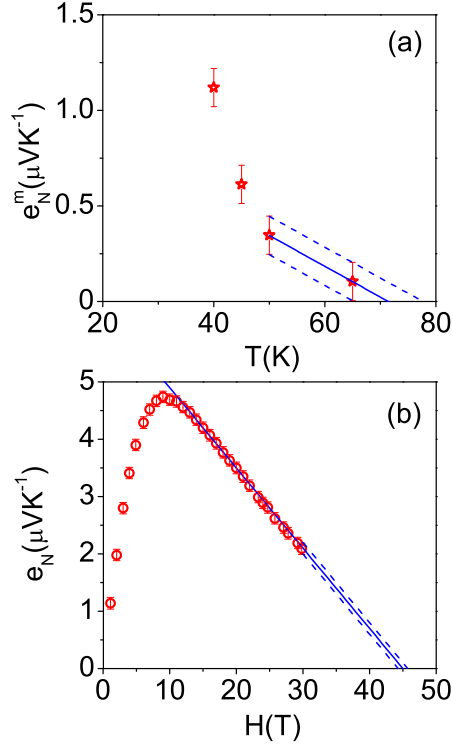}\\
  \caption{Linear extrapolation to determine ${T_v}$ and ${H_{c2}}$ in OP Bi-2201. (a) Linear extrapolations ( blue solid line) of $e_N^m$ at 50 and 60 K to determine ${T_v}$. The blue dashed line represents the error
  bar of extrapolation.  (b) Linear extrapolations ( blue solid line) of $e_N$ in high fields to determine
   ${H_{c2}} (T = 20$ K).}
  \label{TvHc2}
\end{figure}

When $B \to {H_{c2}}$, a linear  dependence of $B$ can be obtained from Eq. (\ref{eN}),
\begin{eqnarray}
e_N \approx {e_T}\frac{{H_{c2}}}{{{B_0} + (H_{c2} + {B_{T}})\exp [t\sqrt {{B_p}/(H_{c2} + {B_{T}})} ]}},
\label{eNtoHc2}
\end{eqnarray}
$H_{c2}$ represents the zero point of the signal. In Fig. \ref{TvHc2}(b), a linear extrapolation of the high fields data, applied to the curve at 20 K, yields $H_{c2}=45\pm0.7$ T. Results of $H_{c2}$ at other temperatures are shown in Table \ref{TDpara}. Since ${B_{l}}\approx H_{c2}'$ ($T_c$=28 K), ${B_{l}}$ can be estimated as $H_{c2}$ (30 K)=48 T.

\begin{table*}[tph]
\begin{center}
\caption{ ${H_{c2}}$ determined  from linear extrapolation of $e_N$ and $e_T$ calculated from the  model $e_T=e_c (T_v/T-1)$.}
\label{TDpara}
\renewcommand\arraystretch{1.5}
\begin{tabularx}{\textwidth}{XXXXXX}
\hline
\hline
$T$ (K)                     &5            &11          & 20           & 30           &40\\
\hline
${H_{c2}}$ (T)              &50.0$\pm$3.3 & 45.0$\pm$1 & 45.0$\pm$0.7 & 48.0$\pm$1.3 &50.0$\pm$3.6\\
\hline
${e_{T}}$ ($\mu$VK$^{-1}$)  &56.8         & 23.5       & 11.0         & 5.91         &3.36\\
\hline
\hline
\end{tabularx}
\end{center}

\end{table*}

At low temperature and fields (i.e., the TAFF regime),  pinning  effect is dominant in damping viscosity.
Thus, $B_0$ can be neglected in $e_N$,
\begin{eqnarray}
{e_N}&\approx&{e_T}\ln(\frac{{{H_{c2}}}}{B})\exp[- ({T_v\over T}-1)\sqrt {\frac{{{B_p}}}{B}}].
\label{eNTAFF}
\end{eqnarray}
Taking the logarithm of the two sides of $e_N$ in Eq. (\ref{eNTAFF}), we can obtain
\begin{eqnarray}
y = {a_1}x + {b_1},
\label{BpfiteN}
\end{eqnarray}
where $y=\ln[\ln(H_{c2}/B)]-\ln {e_N}$, ${a_1}=(T_v/ T-1)\sqrt{{B_p}}$, ${b_1}=-\ln{e_T}$, $x={B^{-1/2}}$. The linear fitting of the data allows us to determine the parameters,
\begin{eqnarray}
{e_T} &=& \exp ( - {b_1}), \nonumber\\
{B_p} &=& a_1^2/{(T_v/ T-1)^2}.
\label{e1Bp}
\end{eqnarray}
The fitting at $T=20$ K is shown in Fig. \ref{eTBp}, and ${B_p}=2.5$ T, ${e_T} = 11.0$ $\mu$VK$^{-1}$ are
obtained. Data in very low fields are not used because the signals are very weak thus the relative error are
big. If a linear decay of superfluid density $n_s=n_{s0}(1-T/T_v)$ is assumed, $e_T=e_c t$ can be obtained at
arbitrary temperature (shown in Table \ref{TDpara}) with $e_c=4.3$ $\mu$VK$^{-1}$.

\begin{figure}
  \centering
  \includegraphics[width=7cm]{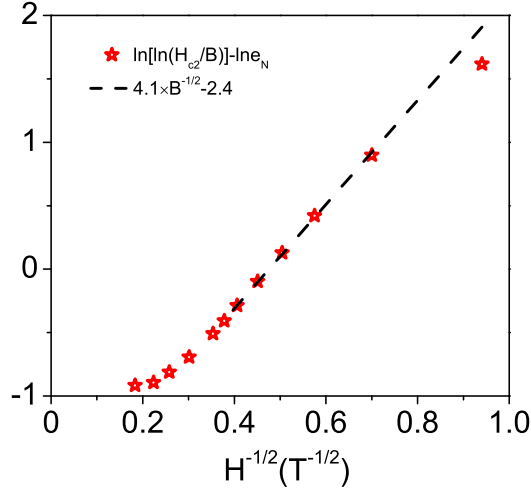}\\
  \caption{Linear fitting of $\ln [\ln (H_{c2}/B)] - \ln {e_N}$ vs ${H^{ - 1/2}}$ to determine ${B_p}$ and
  ${e_T}$  at $T=20$ K.}
  \label{eTBp}
\end{figure}

A nearly linear equation can be obtained from the model Eq. (\ref{eN}) of the Nernst effect at the minimum of $f$ and below $T_c$,
\begin{eqnarray}
g={B\over{B_0+f}},
\label{g}
\end{eqnarray}
where $g=e_N/[e_T\ln(H_{c2}/B)]$. Similar to Eq. (\ref{rhoHTc}), the data in low fields and high temperature
should be used to determine $B_0$. In Fig. \ref{B0b}(a), $g$ vs $B$ at $T=20$ K shows a linear regime in
low fields (from 1 to 8 T), which supports a linear fit ($g=0.032*H$). But the minimum of $f$ is so big (31.1
T) that $B_0$ is found to be very small (0.1 T). In addition, the error of the slope yields that $B_0$ varies
in $\pm1$ T. Thus, $B_0$ can be reasonably chosen as 0.

\begin{figure}
  \centering
  \includegraphics[width=6cm]{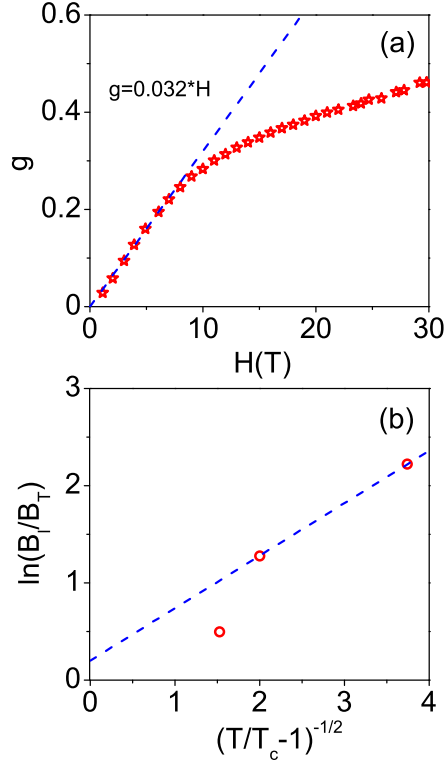}\\
  \caption{Linear fitting to determine $B_0$ and $b$ in OP Bi-2201. (a) Linear fitting($g=0.032*H$, blue solid line) at the minimum of $f$ (Eq. (\ref{f}), red symbols) vs $H$. (b) Linear fitting (blue dashed line) of
  $\ln(B_l/B_T)$ vs $(T/T_c-1)^{-1/2}$ near $T_c=28$ K at 30 K and 35 K.}
  \label{B0b}
\end{figure}

Since ${B_{l}} \approx 48$ T is estimated, $b$ can be determined from the peak value $e_N^m$ above $T_c$, where
\begin{eqnarray}
e_N^m = {e_T}\frac{{B^m}\ln(H_{c2}/{B^m})}{{{B_0} + ({B^m} + {B_{T}})\exp (t\sqrt {\frac{B_p}{B^m +B_{T}}} )}}.
\label{eNm}
\end{eqnarray}
$B_{T}$ can be calculated from this formula directly. According to Korsterlitz model Eq. (\ref{BT}), a linear
model can be obtained as
\begin{eqnarray}
y=a_2x+b_2,
\label{bfiteN}
\end{eqnarray}
where $y=\ln(B_l/B_T)$, $a_2=2b$, $x=(T/T_c-1)^{-1/2}$ and $b_2$ contains the errors. $b$ can be determined from the linear fit of Eq. (\ref{bfiteN}). In Fig. \ref{B0b}(b), the data near $T_c$ (30 K and 35 K) are used to conduct the linear fitting and $b=0.27$ is obtained.

The results of rough estimate are shown in Table \ref{ParaNernst}. Using these results, the comparison between
theory and experiment is carried out and shown in Fig. \ref{roughfit}. They agree with each other, and only
some errors exists in high fields at $T = 20$ K and $T = 30$ K due to the overestimate of $n_s$ with the
linear model $n_s=n_{s0}(1-T/T_v)$.

\begin{figure}
  \centering
  \includegraphics[width=7cm]{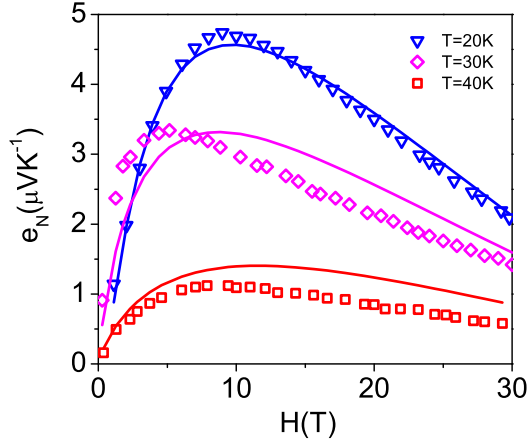}\\
  \caption{Comparison between predictions (solid lines) and experimental data (symbols) using roughly
  determined parameters in Table \ref{ParaNernst}.}
  \label{roughfit}
\end{figure}

\subsubsection{Fine tuning}

Since both $e_T$ and $H_{c2}$ in Eq. (\ref{eN}) are $T$-dependent, the FT procedure of the Nernst effect is a
little more complicated, which requires to run an iteration program. First, the parameters are divided into
two groups. One group is made up of intrinsic parameters $B_0$, $B_p$, $B_{l}$, $b$ and $T_v$.  The other
contains the $T$-dependent parameters ${e_T}$ and $H_{c2}$. The RE helps to define the ranges of the fitting
parameters, e.g., $T_v\in[65.3, 77.7]$ K, $H_{c2}\in[44, 53.6]$ T, $B_l\in[46.7, 49.3]$ T and $B_0=0$.
Besides, $B_p>0$, $b>0$ and $e_T>0$ are set to be constraints. Second, we treat the empirical data of 5 curves at $T$ =5, 11, 20, 30 and 40 K for $H\leq30$ T as an ensemble. Using the iteration program, the FT can be
conducted, one group after another, with the former results substituted into the later calculation. This
iterative process runs until a local minimum (RMSE $\approx0.0993$ $\mu$VK$^{-1}$) in the parameter space
specified as above is reached.

In the Nernst experiment \cite{Wang2006}, the error of the voltage measured by nanovoltmeter is $\pm$5 nV,
thus the uncertainty of the Nernst signal  $\delta e_N=\delta E/\nabla T$ is estimated to be $\pm$25
nVK$^{-1}$. By private communication, Y. Wang indicates that the noise in strong fields should be larger, thus $\delta e_N$ is larger than $\pm$50 nVK$^{-1}$ (the maximum value may be $\pm$0.1 $\mu$VK$^{-1}$ at 45 T),
consistent with the typical fluctuations of the data. In the present work, only data below 30T are used. We
choose a rational fitting error of $\pm$100 nVK$^{-1}$ as the upper boundary of the error function for the
samples except for UD Bi-2201. For UD Bi-2201, the local minimum error is 0.152 $\mu$VK$^{-1}$, thus we set
0.2 $\mu$VK$^{-1}$ as the upper boundary, to reveal the sensitivity of the model. The procedure to determine
the error bar is similar to that above for $\rho$, except $B_l$ is assumed as equals to $H_{c2}$ at $T_c$
(with error bar). The final results of FT are  shown in Table \ref{ParaNernst} and the multi-curves fitting
figure is shown in Fig. \ref{eNcomp}(b).

\end{appendix}

\bibliography{quantitative-nernst-model}

\end{document}